\documentclass[9pt,twocolumn,twoside]{pnas-new}

\templatetype{pnasresearcharticle} 

\makeatletter
\newcommand{\phantomlabel}[2]{
    \protected@write\@auxout{}{
        \string\newlabel{#2}{
            {\@currentlabel#1}{\thepage}
            {\@currentlabel#1}{#2}{}
        }
    }
    \hypertarget{#2}{}
}
\makeatother

\title{Interdependent Diffusion: The social contagion of interacting beliefs}

\author[a,b]{James Houghton}
\affil[a]{Massachusetts Institute of Technology}
\affil[b]{University of Pennsylvania}

\leadauthor{Houghton}

\significancestatement{When studying the spread of beliefs from person to person, scholars generally look at only one belief at a time. This paper shows that by focusing on a single belief, we miss interactions between beliefs that lead to polarization. Because polarization is such an important part of why we study social contagion in the first place, recognizing this oversight encourages us to reexamine much of what we know, and opens up a rich sub-field for future research.}

\correspondingauthor{Please direct correspondence to \href{mailto:houghton@mit.edu}{houghton@mit.edu}.\\ See \href{https://github.com/JamesPHoughton/interdependent-diffusion}{https://github.com/JamesPHoughton/interdependent-diffusion} for simulation code, experimental materials, and data. See \href{https://osf.io/239ns}{https://osf.io/239ns} for experiment preregistration.}

\keywords{Social Contagion $|$ Social Learning $|$ Simulation $|$ Experiment}

\begin{abstract}
Social contagion is the process in which people adopt a belief, idea, or practice from a neighbor and pass it along to someone else. For over 100 years, scholars of social contagion have almost exclusively made the same implicit assumption: that only one belief, idea, or practice spreads through the population at a time \cite{granovetter1973strength,burt1992structural,watts1998collective,watts2011simple, reagans2003network,kempe2003maximizing,lazer2007network,travers2011experimental,kearns2006experimental, acemoglu2011opinion,suri2011cooperation,le1897crowd,degroot1974reaching,schelling1978sorting,granovetter1978threshold,centola2007complex,salganik2006experimental,lorenz2011social,muchnik2013social,shalizi2011homophily,golub2012homophily,christakis2013social,rand2011dynamic,almaatouq2020adaptive,galam1991towards,hegselmann2002opinion}. It is a default assumption that we don’t bother to state, let alone justify. The assumption is so ingrained that our literature doesn’t even have a word for “whatever is to be diffused” \cite{granovetter1973strength}, because we have never needed to discuss more than one of them.
But this assumption is obviously false. Millions of beliefs, ideas, and practices (let’s call them ``diffusants'') spread through social contagion every day. To assume that diffusants spread one at a time – or more generously, that they spread independently of one another – is to assume that interactions between diffusants have no influence on adoption patterns. This could be true, or it could be wildly off the mark. We’ve never stopped to find out.
This paper makes a direct comparison between the spread of independent and interdependent beliefs using simulations, observational data, and a 2400-subject laboratory experiment. I find that in assuming independence between diffusants, scholars have overlooked social processes that fundamentally change the outcomes of social contagion. Interdependence between beliefs generates polarization, irrespective of social network structure, homophily, demographics, politics, or any other commonly cited cause. It also coordinates structures of beliefs that can have both internal justification and social support without any grounding in external truth.
\end{abstract}

\dates{This manuscript was compiled on \today}

\begin{document}

\maketitle
\thispagestyle{firststyle}
\ifthenelse{\boolean{shortarticle}}{\ifthenelse{\boolean{singlecolumn}}{\abscontentformatted}{\abscontent}}{}

\dropcap{O}n June 17\textsuperscript{th}, 2015, the U.S. Treasury announced that a portrait of a woman would appear on the ten-dollar bill \cite{calmes_2015}. The same day, news emerged of a mass shooting at a historically black church in Charleston, South Carolina \cite{houghton2019beyond,hawes2019grace}. Within twenty-four hours both stories had spread throughout U.S. social media. It is reasonable to study the spread of these diffusants independently of one another, because the probability that an individual will share news of the shooting is not likely to be causally influenced by whether they have understood news about the \$10 bill, and vice versa.

The day after the Charleston shooting, two other news items emerged. The first was a report that the shooter had been motivated by racial hatred symbolized in the Confederate flag. The second was a call to remove the Confederate flag from the South Carolina state capitol grounds \cite{houghton2019beyond,hawes2019grace}. Even though these are distinct ideas, each spreading through social contagion, we cannot ignore one in trying to understand the diffusion of the other. If an individual has previously adopted the belief that the flag should be removed from the capitol grounds, they will be more likely to believe that the shooter’s identification with the flag is politically relevant, and vice versa. Rather than being independent diffusants, these beliefs are interdependent.

With good reason, nearly all social contagion research assumes that diffusants spread independently of one another. The independence assumption makes for parsimonious theory \cite{granovetter1973strength, burt1992structural, watts1998collective, watts2011simple, reagans2003network, kempe2003maximizing, lazer2007network, le1897crowd, degroot1974reaching, schelling1978sorting, granovetter1978threshold, centola2007complex, shalizi2011homophily, golub2012homophily, christakis2013social,galam1991towards,hegselmann2002opinion,acemoglu2011opinion}, and it reduces the complexity and expense of experiments \cite{travers2011experimental, kearns2006experimental, suri2011cooperation, salganik2006experimental, lorenz2011social,muchnik2013social, rand2011dynamic,almaatouq2020adaptive}. The most influential authors on social contagion have used this assumption to study the effect of social network structure \cite{granovetter1973strength,burt1992structural,watts1998collective,watts2011simple, reagans2003network,kempe2003maximizing,lazer2007network,travers2011experimental,kearns2006experimental,suri2011cooperation,acemoglu2011opinion}, social reinforcement \cite{le1897crowd,degroot1974reaching,schelling1978sorting,granovetter1978threshold,centola2007complex,salganik2006experimental,lorenz2011social,muchnik2013social,hegselmann2002opinion}, homophily \cite{shalizi2011homophily,golub2012homophily,christakis2013social}, and network rewiring \cite{christakis2013social,rand2011dynamic,almaatouq2020adaptive} on contagion outcomes. Unfortunately, scholars assume independence between diffusants so frequently and to such productive ends that we generally forget we are doing so, and fail to question whether the assumption is appropriate.

In contrast with independent diffusion, “interdependent diffusion” describes any social contagion process in which individuals’ likelihood of adopting diffusant $A$ is a function of their current state of adoption of $B$ ($C$, $D$, $\ldots$) and in which their likelihood of adopting $B$ ($C$, $D$, $\ldots$) is a function of their state of adoption of $A$. Very few theoretical models include any form of interaction between diffusants \cite{baldassarri2007dynamics,dellaposta2015liberals, friedkin2016network, goldberg2018beyond, xiong2017analysis, axelrod-1997-dissemination}, and none are empirically verified. It remains to be seen whether interdependence between diffusants 1) generates new sociological processes, 2) creates new observable outcomes, and 3) has practical consequences for communication and social policy. In short, does interdependence matter for the theoretical and empirical study of social contagion, or are our models of independent diffusion sufficient?

\section*{Theoretical development}
Interdependent diffusion can be studied through two theoretical lenses. First, we will observe the spread of a ``focal'' belief as it participates in a process of ``reciprocal facilitation'' with other diffusants. Secondly, we will observe a pair of individuals as they exchange multiple beliefs with one another in a process we might call an ``agreement cascade''.

\begin{figure*}[ht]
    \includegraphics[width=7in]{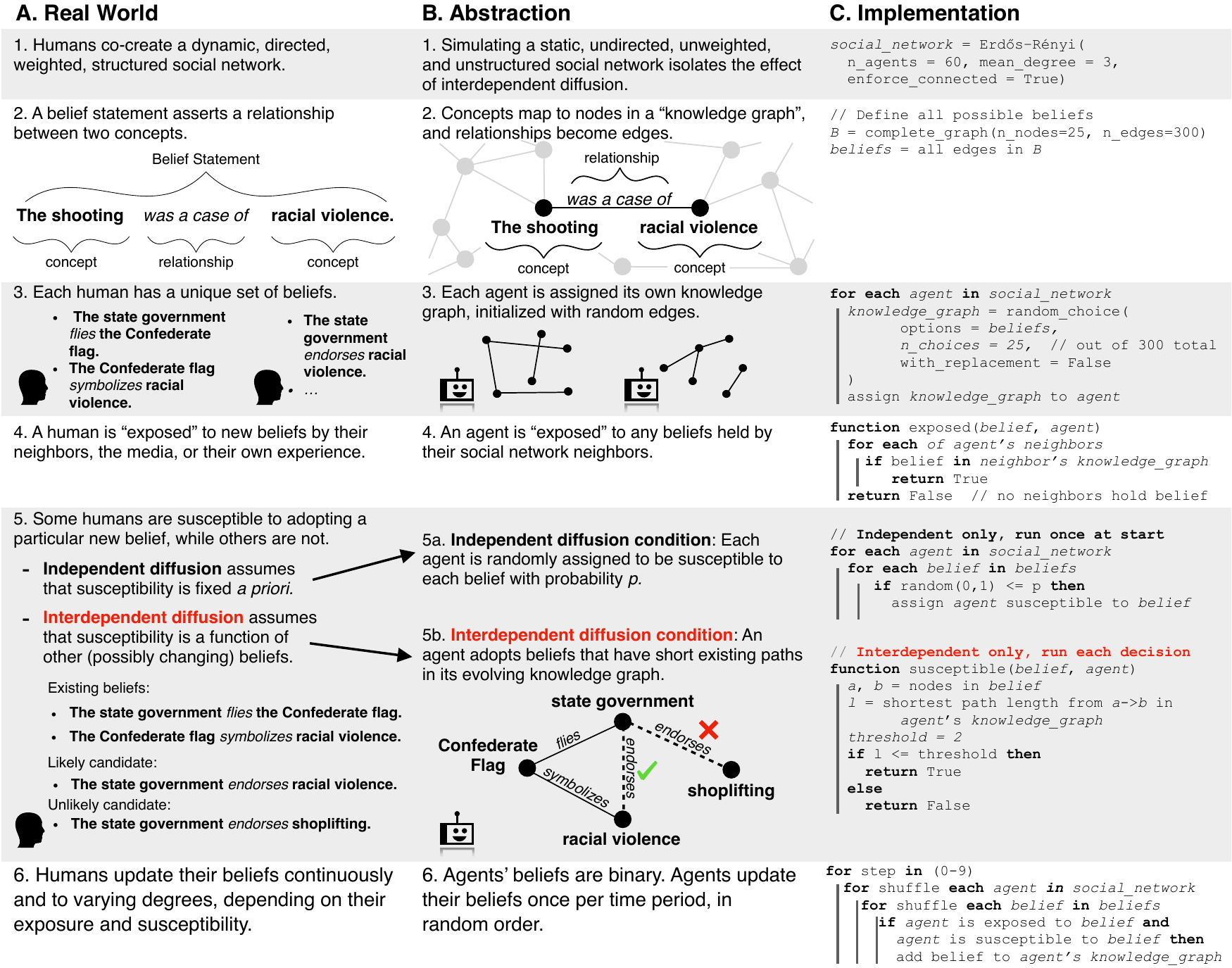}
    \caption{\textbf{Simulating interdependent diffusion as knowledge graph edges spreading over a social network.} \textbf{A.} In the real world, the effect of interdependence on social contagion is enmeshed with other influences and complexities.  \textbf{B.} A useful abstraction simplifies reality to isolate the mechanisms of interest, and \textbf{C.} allows us to simulate the behavior we wish to understand.}
    \label{fig:model}
\end{figure*}

\subsection*{Reciprocal Facilitation}
A focal belief spreads when an exposed individual’s existing beliefs lead her to think it is true – that is, they ``facilitate'' her adoption of the focal belief. From its new position in the social network, the focal belief may then spread further, and subsequently facilitate the adoption of the beliefs that had previously supported its diffusion. This cycle creates a reinforcing feedback, such that when diffusants alternately create susceptibility to one another they can be adopted by more individuals than any single belief could have reached on its own. We see this ``reciprocal facilitation'' dynamic at work in beliefs about the confederate flag. The existing political conflict facilitated the spread of information about the shooter’s identification with the flag, and as it spread, news of the shooter’s identification with the flag brought more attention to its contentious display at the state capitol.

We can use a simple agent-based simulation (described in Fig. \ref{fig:model}) to explore the macro-scale effects of reciprocal facilitation.
In this simulation, agents populate a fixed, random social network, and are initially assigned a random set of beliefs. Each belief asserts a relationship between two concepts (\textit{e.g.} \textbf{the shooting} \textit{was a case of} \textbf{racial violence}). By analogy to the ``knowledge graph'' interpretation of semantic networks\cite{sowa1987semantic, collins1975spreading, popping2003knowledge}, an agent's beliefs can be aggregated to form a graph in which nodes represent concepts (\textit{e.g.} \textbf{people}, \textbf{places}, \textbf{items}, \textbf{activities}, etc\ldots), and edges represent relationships that connect concepts to one another (\textit{e.g.} \textit{ownership}, \textit{membership}, \textit{co-location}, \textit{usage}, etc\ldots). When an agent adopts a new belief, it becomes a new edge in the agent's knowledge graph.

When an individual is exposed to new beliefs, she prefers to adopt those that connect two concepts that are already close together in her knowledge graph \cite{schilling2005small,nickerson1998confirmation}, as doing so does not require dramatic changes to her belief structure. The simplest representation of this tendency is that a simulated individual will adopt any edge present in at least one neighbor's knowledge graph, as long as the distance it spans in her existing knowledge graph is below some threshold. This decision rule ignores social reinforcement (\textit{i.e.} no complex contagion) and the exposer's attributes or other beliefs (\textit{i.e.} no homophily or heterogeneous influence), and so for each belief is equivalent to a simple contagion model in all respects other than the influence of the individual's internal state. Please see the \hyperref[methods]{\textbf{Methods}} section and section S1 of the supplement for simulation details.

Figure \ref{fig:sim1} shows the average results of 20,000 simulations of 60 agents in the above model. Each agent is assigned random initial beliefs, and adopts beliefs from its social network neighbors that span up to two steps distance in its knowledge graph. First we see that under interdependent diffusion, reciprocal facilitation allows the average number of people susceptible to a belief to grow endogenously with the number of adopters (Fig. \ref{fig:sim1:A}, red). By comparison, if the same beliefs diffuse independently they can only be adopted by individuals who are susceptible at the start, and so they spread less widely in the population (Fig. \ref{fig:sim1:A}, black). Put another way: to explain the same level of final adoption, models of independent diffusion need to assume that significantly more people start out susceptible to each belief.

It might be reasonable for us to ignore interdependence (and instead assume more widespread initial susceptibility) were it not for the second effect of reciprocal facilitation. When diffusants do not interact, the number of people initially susceptible to a belief is an excellent predictor of how many will eventually adopt it (Fig. \ref{fig:sim1:B}). However, when beliefs are interdependent, individuals can acquire susceptibility by adopting other supporting beliefs, such that initial susceptibility is no longer a strong predictor of eventual adoption. For example, imagine that a “focus group” is selected from our artificial population before the simulation starts, and that amongst this group belief $A$ is adopted by 25\% more people than belief $B$. Unsurprisingly, if we simulate the spread of these beliefs independently of one another, belief $A$ will be more popular than $B$ in over 99\% of cases. On the other hand, when we simulate interdependent diffusion, $A$ is adopted by more people than $B$ only 57\% of the time – just slightly better than chance.

\begin{figure}[ht]
    \includegraphics[width=\linewidth]{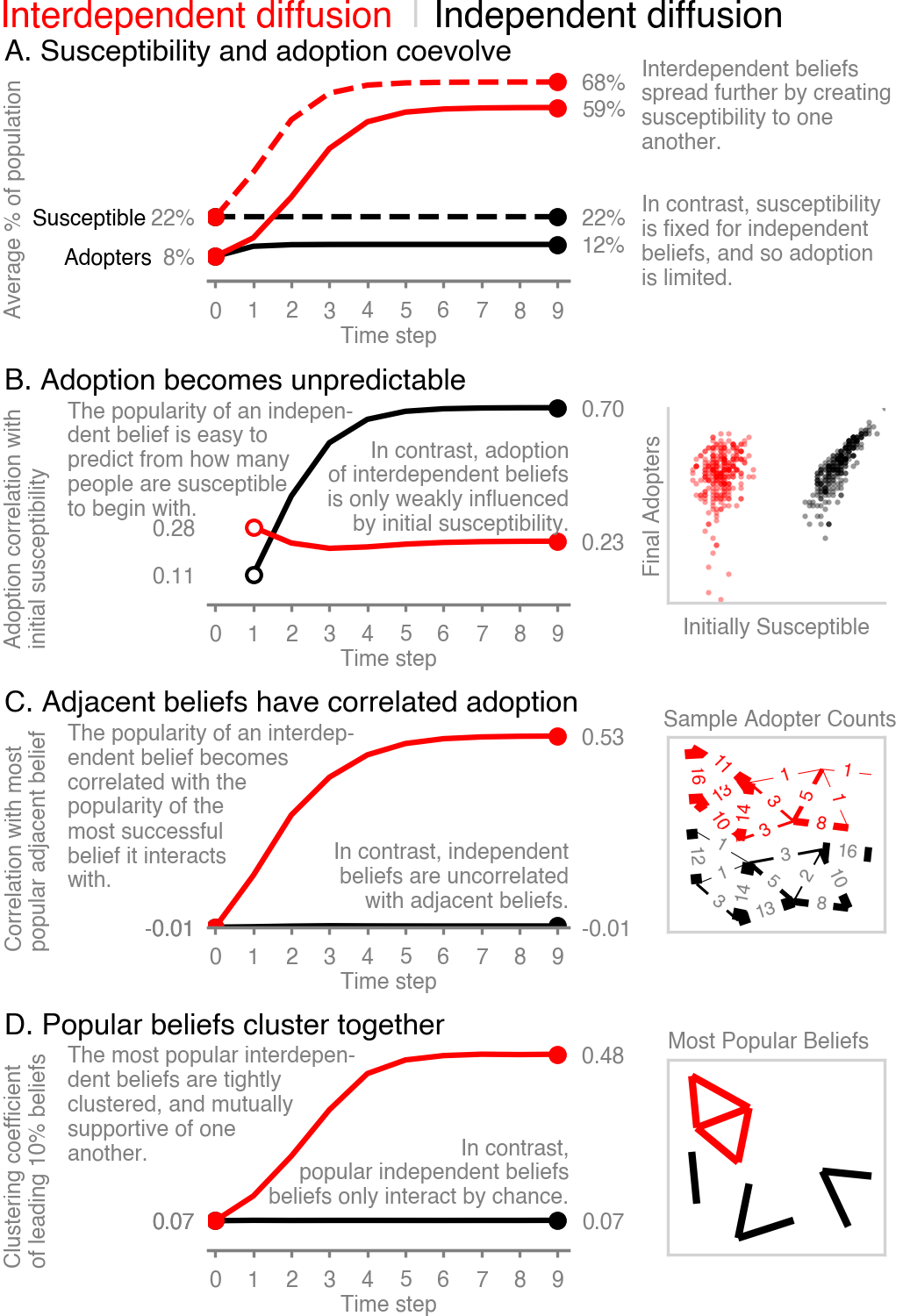}
    \caption{\textbf{Simulated effect of interdependent and independent diffusion on shared belief structures.} \textbf{A.} Average popularity (number of adopters) of all beliefs over time, given an equal number of initially susceptible individuals in each condition. \textbf{B.} Correlation between the initial number of susceptible individuals and the popularity of each belief, given the same \textit{final} distribution of adoption between conditions; detail shows $t=9$ popularity vs. number of individuals initially susceptible. \textbf{C.} Correlation between the popularity of each belief and the popularity of the most popular belief it shares a 'node' with; sample $t=9$ popularity of each belief overlaid on the knowledge graph. \textbf{D.} Clustering coefficient of a hypothetical knowledge graph made up of the most popular 10\% of all beliefs; sample $t=9$ network filtered on popularity.}
    \label{fig:sim1}
    \phantomlabel{A}{fig:sim1:A}
    \phantomlabel{B}{fig:sim1:B}
    \phantomlabel{C}{fig:sim1:C}
    \phantomlabel{D}{fig:sim1:D}
\end{figure}

The popularity of interdependent beliefs is hard to predict because the reciprocal facilitation process does not support the diffusion of all beliefs to the same degree. Instead, it preferentially distributes those that are supported by other widely adopted beliefs. Fig. \ref{fig:sim1:C} shows how this reinforcing feedback leads a belief’s popularity to become correlated with that of the most popular belief adjacent to it in the knowledge graph. This is because a belief that is supported by many popular beliefs will have many opportunities to diffuse, and then to facilitate the adoption of other closely related beliefs. Conversely, a belief that is only supported by unpopular beliefs will have trouble reaching even the few individuals who are susceptible to adopting it.

As a result, patterns emerge when individuals' knowledge graphs are aggregated to the level of the population. After interdependent diffusion, clusters of mutually-supporting beliefs are held by large fractions of the population (Fig. \ref{fig:sim1:D}), and they dictate which new beliefs can be adopted. This is quite remarkable, as \textit{post-hoc}, individuals have both internal support for their beliefs (\textit{i.e.} each belief is supported by many other beliefs) and external support (\textit{i.e.} other individuals share both their beliefs and their justifications for believing so). This happens even though \textit{a priori} we have no way to tell which sets of beliefs will become popular, and no ground truth.

Reciprocal facilitation yields an emergent macro-scale outcome analogous to the micro-scale effect of confirmation bias, in which whole groups of people can collectively deceive themselves through the rational action of comparing new beliefs for consistency with what they already know. This effect doesn't require agents to have any prior preference for believing one thing over another, nor does it depend on any in-group vs. out-group sentiment, affinity for similar individuals, network structure, or social reinforcement. It's merely a phenomenon of the social contagion of interdependent beliefs.

\subsection*{Agreement Cascades}

\begin{figure}[h]
    \includegraphics[width=\linewidth]{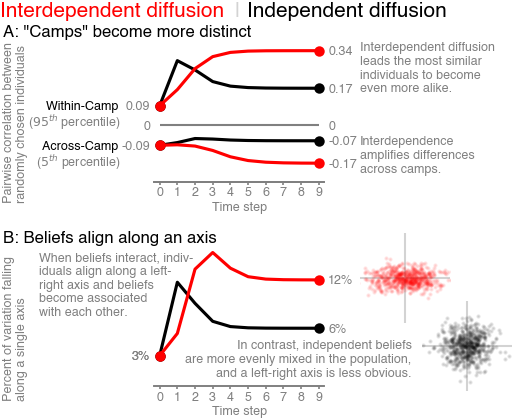}
    \caption{\textbf{Simulated effect of interdependent and independent diffusion on social polarization.} \textbf{A.} Correlation between beliefs of randomly sampled pairs of individuals, $95^{th}$ ($5^{th}$) percentile represents relationships conservatively within (across) ideological "camps". \textbf{B.} An individual's beliefs locate her in a space with one dimension for each possible belief (300). The percent of variation explained by first principal component of this space describes how well individuals map to a "political spectrum". Offset diagrams are exaggerated and show a larger population to illustrate how a component can explain more or less variation. }
    \label{fig:sim2}
    \phantomlabel{A}{fig:sim2:A}
    \phantomlabel{B}{fig:sim2:B}
\end{figure}

Switching lenses, we now focus on a pair of neighboring individuals, and observe a process of “agreement cascades”. When two people exchange beliefs, they become more similar to one another. Because their existing beliefs influence the way people respond to new diffusants, shared beliefs make the two individuals more likely to adopt (or reject) the same new beliefs in the future. As a result, they become more similar still, regardless of any preference to align or distinguish themselves from one another.

It is helpful to contrast agreement cascades with homophily (\textit{i.e.} an individual's tendency to give extra weight to the opinions of more similar neighbors \cite{axelrod-1997-dissemination}). When homophily is active, an individual is (consciously or unconsciously) aware of their similarity to a neighbor, and allows that assessment of similarity to influence their adoption decisions. In contrast, agreement cascades increase an individual's likelihood of adopting a belief from a similar neighbor, even if she is blind to every other belief her neighbor possesses. Homophily increases the attractiveness of all beliefs from a similar neighbor, while the same belief shared by a dissimilar neighbor would be less attractive. By contrast, under interdependent diffusion each belief has a particular likelihood of adoption regardless of its source, but agreement cascades mean that similar neighbors provide a higher concentration of beliefs that an individual is likely to adopt.

Fig. \ref{fig:sim2:A} shows the similarity between pairs of highly similar individuals (whom we might describe as being in the same ``ideological camp'') and highly dissimilar individuals (different camps). Under interdependent diffusion, agreement cascades lead camps to become increasingly self-similar as they expand members’ access to beliefs they can hold in common, and filter out beliefs that would set them apart from one another. Moreover, differences between camps are amplified as existing beliefs drive dissimilar individuals to adopt beliefs from different parts of the belief space.

As individuals organize into camps, it becomes easier to predict each person’s position on one belief from their position on other beliefs. For example, belief $A$ may co-occur with belief $B$ 70\% of the time, but co-occur with belief $C$ only 10\% of the time. Interdependent diffusion compresses the population’s variation in the space of beliefs into a few principal axes (\textit{e.g.} liberal-conservative, or libertarian-populist). As a result, individuals can be more easily described by their position along a “left-right axis”, as seen in Fig. \ref{fig:sim2:B}, with individuals at any point on the axis relatively similar to one another.

\section*{Consistency with observational data}
If reciprocal facilitation and agreement cascades are active in the real world, we should be able to observe the simulation’s predictions in real-world data. For example, when a document asserts a connection between two concepts, then the most popular connections drawn from a corpus of related documents should exhibit more clustering than we expect by chance. Figure \ref{fig:observational:A} shows this effect in article keywords from the New York Times \cite{gallina-etal-2019-kptimes} and academic paper keywords listed in the Web of Science \cite{kowsari2017HDLTex}, over the full range of possible thresholds for a ``popular'' keyword co-occurrence.

The simulation also predicts that the most similar individuals in the population should be more alike than expected by chance, and those who are dissimilar more different. Figure \ref{fig:observational:B} shows that this prediction is borne out in surveys of political opinions by the Pew Research Center \cite{pew2014} and in surveys of social values by the World Values Survey \cite{wvsa2020}.

Finally, the simulation predicts that the variation between individuals in these surveys should be more aligned with a few principle axes of polarization than we expect by chance. Figure \ref{fig:observational:C} shows this to be true. Please see the \hyperref[methods]{\textbf{Methods}} section and section S2 of the supplement for analysis details.

\begin{figure}[hb]
    \includegraphics[width=\linewidth]{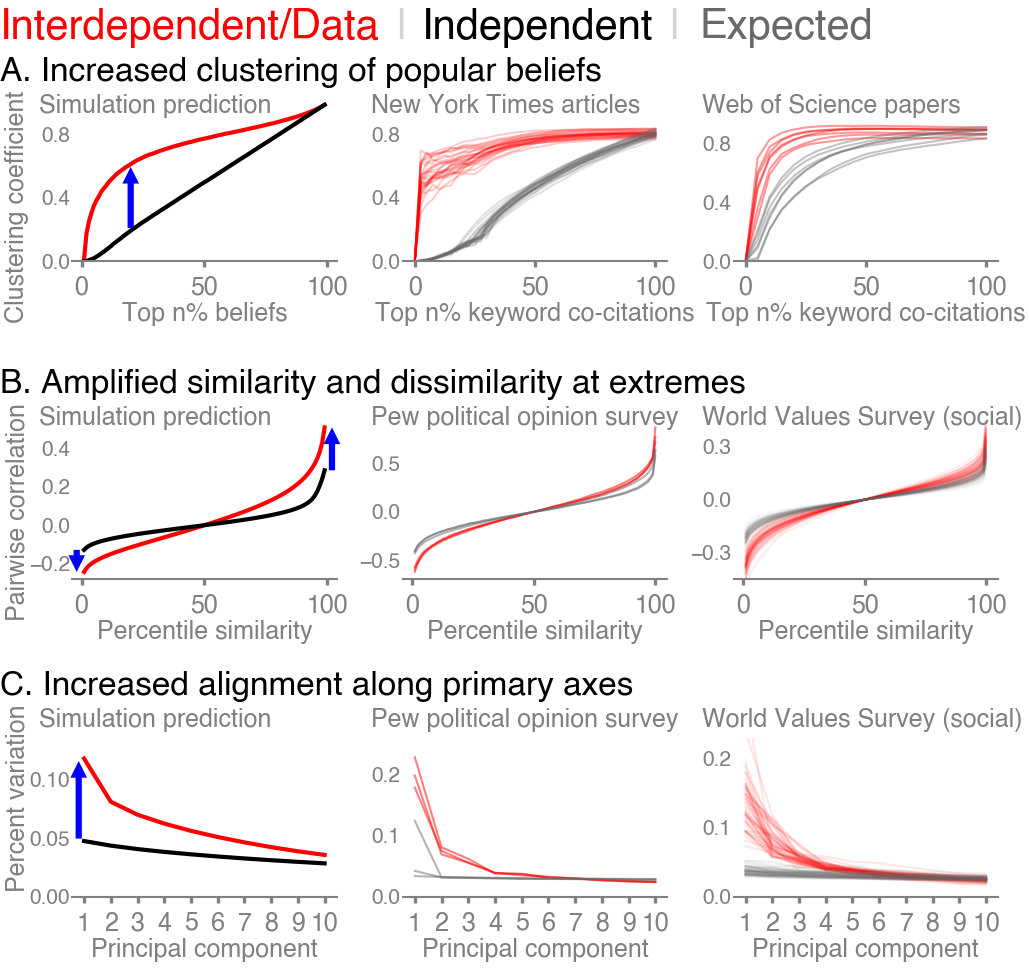}
    \caption{\textbf{Observing the predicted effects of interdependent diffusion in real-world data.} A. The clustering coefficient of a hypothetical knowledge graph constructed from the $n\%$ most popular beliefs, for $n\in(0,100)$. Interdependence is predicted to increase clustering among the most popular connections between concepts. This increase is observed in keyword co-citation networks from 289k New York Times articles (curves for each of 27 sections)\cite{gallina-etal-2019-kptimes}, and 47k academic papers from Web Of Science (7 fields)\cite{kowsari2017HDLTex}, compared to a randomized baseline.  B. Correlation in beliefs between randomly selected pairs of individuals. Interdependence is predicted to amplify extreme values of similarity and dissimilarity. This amplification is observed among 10k responses to the Pew Political Polarization Survey (3 waves)\cite{pew2014}, and 71k responses to questions about social values in the World Values Survey (49 countries)\cite{wvsa2020}. C. The percentage of variation present in the first 10 components of a principal component analysis. Interdependence is predicted to increase the variance explained by the first few components. This increase is observed in the Pew and World Values surveys.}
    \label{fig:observational}
    \phantomlabel{A}{fig:observational:A}
    \phantomlabel{B}{fig:observational:B}
    \phantomlabel{C}{fig:observational:C}
\end{figure}

\section*{Testing predictions of population-level outcomes in a randomized controlled experiment}
The above simulation describes two mechanisms involved in the social contagion of interacting beliefs, and predicts outcomes that are observed in real world data. However, this model is a simplification of reality, and like all such models may mischaracterize human behavior in important ways. Similarly, observational analysis is limited in this case by the lack of an independent counterfactual; it can only show consistency with the predicted outcomes, rather than causal evidence of the mechanisms themselves. To rigorously test the simulation’s predictions, I conducted a fully preregistered randomized-controlled experiment with 2400 participants. In an online laboratory context, I systematically varied the level of interaction between otherwise identical diffusants in identically constructed populations, and measured macro-scale polarization in behavior and self-reported beliefs.

The experiment took the form of a ``detective game'' in which participants were given clues to a burglary and were asked to sort those clues into ``Promising Leads'' and ``Dead Ends'', using the interface in Fig. \ref{fig:game:A}.  When a participant categorized a clue as a promising lead, it was immediately shared with her three neighbors in a 20-person social network, and so could diffuse through the network.

In our simulation model, agents could be programmed to either pay attention to interactions between beliefs or to ignore them.
Human beings, on the other hand, are wired to see connections between ideas, and so it is impossible to conduct a perfect test in which all clues interact in one experimental condition, and yet \textit{the same clues} are fully independent in another. To approximate the ideal manipulation, I partitioned the clues such that some (22 clues) were common to both interdependent and independent conditions and were used for analysis, while the remainder (55 clues) varied across conditions to manipulate the level of interaction between analysis clues. In both conditions, analysis clues linked a crime scene and stolen object to three suspects, two articles of clothing, two physical descriptions, two tools, and two vehicles (Fig. \ref{fig:game:B}). In the interdependent condition, ``cross-link'' clues connected each suspect, vehicle, etc. to one another (Fig. \ref{fig:game:C}), to create a fully-connected knowledge graph. In the independent condition, ``filler'' clues gave additional, non-interacting details (Fig. \ref{fig:game:D}) that allowed analysis clues to remain as independent as possible. Clues were designed to equally implicate any suspect or burglary method, and each was randomly assigned to exactly one player at game start.

The experiment was run in 30 blocks of matched treatment/control games, each block using a unique set of randomly generated clues. Each social network position was randomly assigned four clues, common (apart from the manipulation) across conditions. Each participant was randomly assigned to a condition and social network position. Groups played for eight minutes before estimating the likelihood that each suspect, vehicle, etc. was involved in the crime.

\begin{figure}[ht]
    \includegraphics[width=\linewidth]{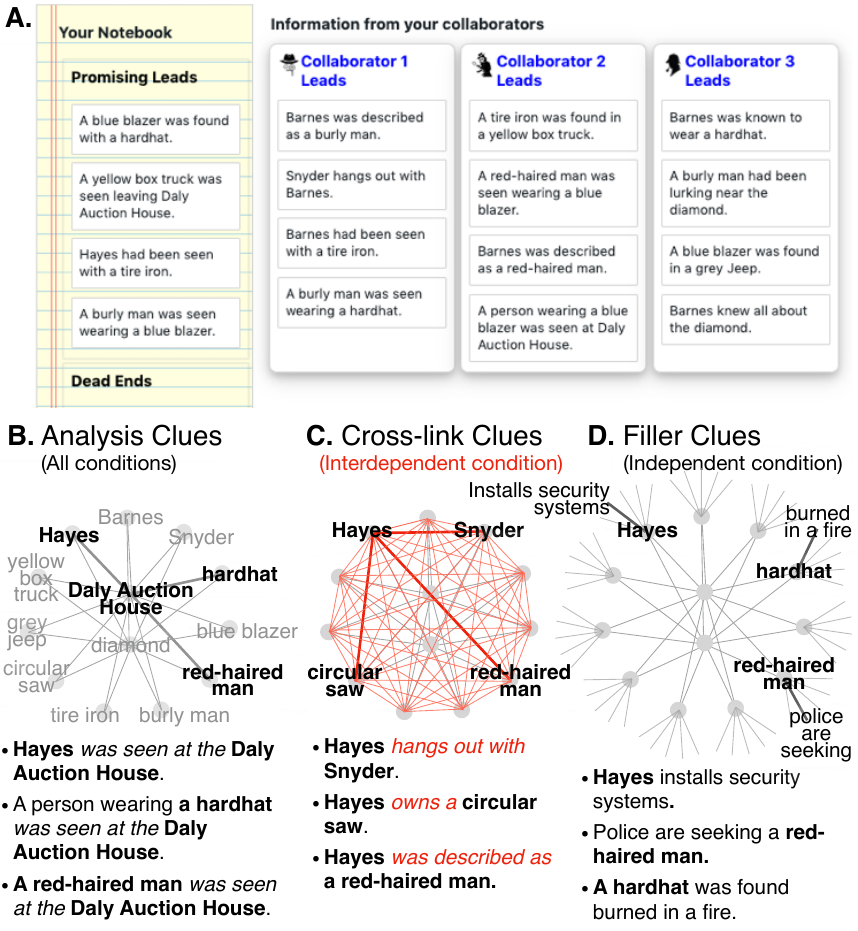}
    \caption{\textbf{The Detective Game.} \textbf{A.} Player Interface. \textit{``A priceless diamond has been stolen from the Daly Auction House. To solve the mystery, drag clues that you think are true into the ``Promising Leads'' section of your notebook, and clues you think are false into the "Dead Ends" section. You are rewarded for each correct 'lead' and penalized for each incorrect 'lead'. There is no reward or penalty for 'dead ends'.''}
    \textbf{B.} 22 ``analysis'' clues link the crime scene and stolen object to three suspects, two articles of clothing, two descriptions, two tools and two vehicles.
    \textbf{C.} In the interdependent condition, 55 ``cross-link'' clues connect each analysis clue to every other analysis clue.
    \textbf{D.} In the independent condition, ``filler clues'' take the place of cross-link clues, allowing analysis clues to remain independent of one another.
    }
    \label{fig:game}
    \phantomlabel{A}{fig:game:A}
    \phantomlabel{B}{fig:game:B}
    \phantomlabel{C}{fig:game:C}
    \phantomlabel{D}{fig:game:D}
\end{figure}

The measures of similarity and left-right axis alignment used in Fig. \ref{fig:sim2}
were computed based on participants’ behavior and self-reported opinions.
``Behavioral'' measures were constructed from each participant’s final categorization of the 22  analysis clues.
“Self-report” measures were constructed from the 11-item post-game assessments, reflecting how participants internalized the information they encountered to create opinions. As our simulation gives a strong theoretical prior for the direction of each effect, I used uncorrected pairwise one-sided t-tests to assess how each measure differed between baseline and treatment conditions.
For more details, please see the \hyperref[methods]{\textbf{Methods}} section and supplement section S3.

In a testament to the pervasiveness of belief interaction, even this manipulation (and ideal laboratory conditions) could not create a control condition with perfectly independent beliefs. For example, even though the independent clue set contained no explicit links between suspects, participants could draw implicit connections between the guilt of one suspect and the presumed innocence of another. This imperfect control means that measured differences between independent and interdependent conditions are likely to underestimate the true effect of belief interaction.

Nevertheless, the results of this experiment support the above macro-level predictions. Figure \ref{fig:experiment:A} shows that interdependence measurably increased the population’s alignment along a left-right axis among both behavioral (+2.1\%, $p=.013$, $90\%~CI=(0.6,3.7)$) and self-reported (+2.8\%, $p=.022$, $CI=(0.6,5.0)$) measures of belief. While not all measures were significant, camps were more self-similar (behavioral measures: +.026, $p=.015$, $CI=(0.008,0.045)$, Fig. \ref{fig:experiment:B}) and more distinct from one another (self-report measures: -.034, $p=.058$, $CI=(-0.07,0.0015)$, Fig. \ref{fig:experiment:C}) in the interdependent condition than in the independent condition.

\subsection*{Comparison effect of social network structure}
To gauge whether the effect of interdependence is large enough to be worth attention when compared to other drivers of polarization, I ran a parallel experimental condition that fixed diffusants to be independent and varied the social network structure between non-polarizing and polarizing extremes. The baseline was a “dodecahedral” network (inset diagram in Fig. \ref{fig:experiment:A}), in which none of a participant’s neighbors were directly connected to any other (network clustering coefficient $0.0$), and the average social network distance between individuals was short ($2.6$ steps). We should expect to find very little polarization in this network, as information can diffuse across the network readily, and coordination among subgroups is impeded by the lack of mutual connections. The second social network was a “regular connected caveman” structure, in which neighbors shared $50\%$ of their remaining contacts in common (clustering coefficient $0.5$), and there are large average distances between individuals ($4$ steps). We should expect to find high levels of polarization in this network regardless of the level of interaction between clues, as strong clustering makes it easy for subgroups to converge on a shared set of clues, and long average path lengths make it harder for information to spread between camps.

\begin{figure}[ht]
    \includegraphics[width=\linewidth]{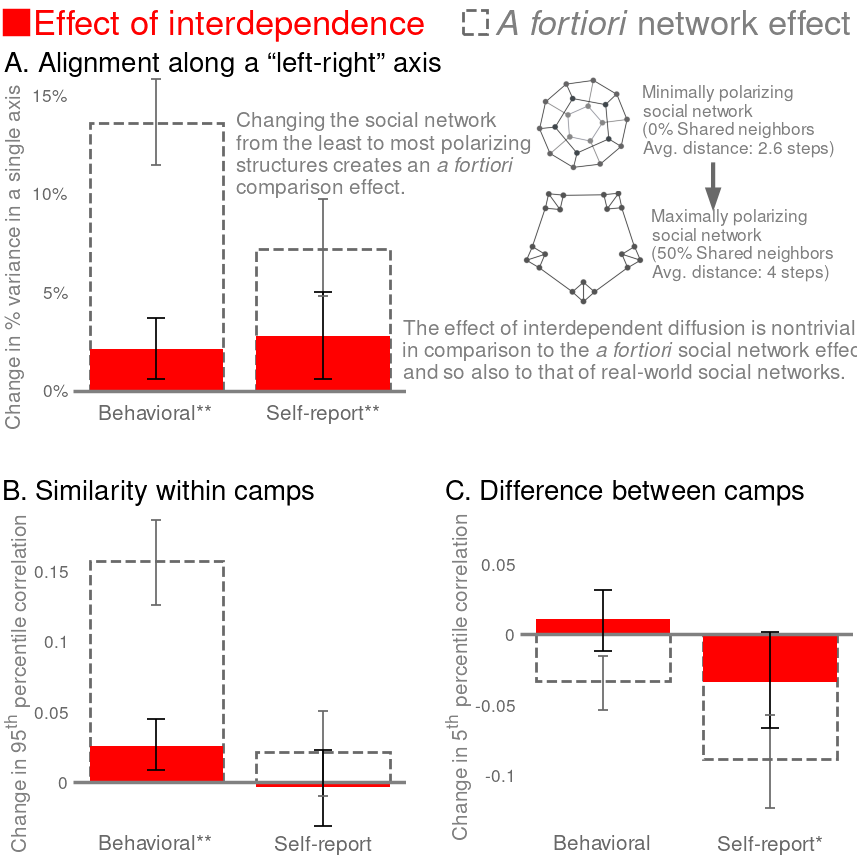}
    \caption{\textbf{Testing the macro-scale effects of interdependent diffusion in an online lab experiment.} The predicted effects of belief interaction are borne out in experiment, with meaningful effects in comparison to an \textit{a fortiori} comparison that varied the social network structure underlying independent diffusion. Paired T-test $^*p<.1$, $^{**}p<.05$. Error bars show 90\% bootstrap CI. $n=30$ pairwise comparisons.}
    \label{fig:experiment}
    \phantomlabel{A}{fig:experiment:A}
    \phantomlabel{B}{fig:experiment:B}
    \phantomlabel{C}{fig:experiment:C}
\end{figure}

Real-world social networks lie somewhere between these two extremes, and so this social network manipulation creates an \textit{a fortiori} comparison effect. If the effect of interdependence is nontrivial in this comparison, then we have confidence that it is meaningful in the real world. In the statistically significant measures of this experiment, interdependent diffusion creates an effect between 16\% and 39\% of the \textit{a fortiori} reference, as shown in Fig. \ref{fig:experiment}. We should always use caution when generalizing effect sizes from laboratory groups to large-scale social networks. However, this comparison suggests that interdependence plays an unignorable role in social contagion, and that scholars’ almost exclusive focus on network structure over belief interaction is out of proportion to the relative importance of the two effects. Please see the suplement section 3.6 for comparison details.

\section*{Discussion}
While it cannot explore all of the implications of interdependent diffusion, this paper demonstrates that belief interaction enables at least two new sociological processes that influence polarization and shared belief structures. The social contagion literature is replete with studies that explain polarization as a consequence of social network structure \cite{flache2011small,del2016echo}, homophily \cite{dandekar2013biased,vasconcelos2019consensus,axelrod-1997-dissemination}, complex contagion \cite{spohr2017fake,tornberg2018echo}, or other personal or network characteristics. While they certainly influence polarization, this paper shows that when beliefs interdepend, none of these explanations are necessary for polarization to emerge - polarization is a natural consequence of diffusion itself. Likewise, we have ample research asking how fringe beliefs and conspiracy theories emerge and persist \cite{spohr2017fake,tornberg2018echo,pennycook2019lazy}. Interdependent diffusion suggests that they are a natural consequence of social contagion.

Given that truly independent diffusion is likely to be rare among real world beliefs, we can no longer assume that consensus and truth should be the default outcome of social contagion, absent other polarizing factors. No matter what changes we make to the social network structure, or how we reduce interpersonal bias, we should still expect polarization to emerge. This motivates several new research questions: What are the conditions under which populations can overcome the polarizing effect of interdependent diffusion to reach consensus? How can populations avoid deceiving themselves about the truth of the world? Answering these questions will require us to drop the assumption of independence and confront the complex reality of belief interaction in social contagion.

This simulation and experiment remind us that the simplifying assumption of independence between diffusants is just that – an assumption. Despite its ubiquity, the assumption should be carefully made and frequently challenged. With luck we will find that most of what we know about diffusion is robust to interaction between diffusants. We may also find that relaxing the assumption of independence helps us explain new sociological phenomenon and better understand social contagion.

\matmethods{
\label{methods}
All materials used to create this paper (including code for running simulations, observational analysis, designing and conducting the experiment, and analyzing experimental results, along with anonymized, timestamped experimental data) is available at:  \href{https://github.com/JamesPHoughton/interdependent-diffusion}{https://github.com/JamesPHoughton/interdependent-diffusion}.

\subsection*{Simulation}
In the simulation presented in \textbf{Figs. \ref{fig:sim1} and \ref{fig:sim2}}, the social network is a connected Erdős–Rényi ($G_{nm}$) random graph with 60 agents, each with an average of 3 neighbors. Each agent is initialized with 25 beliefs (edges) selected randomly from the 300 edges available in a complete knowledge graph with 25 concepts (nodes). These values ensure good coverage of beliefs in the population, while individual knowledge graphs are initially sparse. Random seeding ensures that the simulation starts without polarization or systematic variation in belief popularity, and also that the social network structure itself does not contribute to polarization. Because beliefs are drawn from a complete knowledge graph, there is no natural belief structure around which polarization may nucleate. Results are qualitatively similar with other types and sizes of social networks and different sizes and seeding densities of initial knowledge graphs, so long as there are enough beliefs seeded initially for diffusion to occur and not so many that adoption is complete.

In each step, individuals are selected in random order and update their beliefs by incorporating into their knowledge graphs all beliefs (edges) that their neighbors possess and they are susceptible to adopting. In the interdependent case, individuals are susceptible to any belief with an existing path length of 2 (\textit{i.e.} that closes a triangle) in their knowledge graphs at the current time. In the comparison (independent) case for \textbf{Fig. \ref{fig:sim1:A}}, a random selection of the population is defined to be susceptible to each belief in the same proportion as are initially susceptible to the belief in the interdependent diffusion condition. In \textbf{Figs. \ref{fig:sim1:B}, \ref{fig:sim1:C}, \ref{fig:sim1:D} and \ref{fig:sim2}}, a random selection of susceptible individuals is made in proportion to the final number of susceptible individuals in the interdependent diffusion case. As a result, for all graphs other than \textbf{Fig. \ref{fig:sim1:A}}, a histogram of the extent of diffusion of each belief is approximately the same under both independent and interdependent treatments. This ensures that the subsequent presentation of results reflects purely the effect of interdependence between diffusants, and not the effect of different levels of adoption in the compared populations. Results are qualitatively similar when calculated based upon the initial susceptibility.

\subsection*{Measures}

In \textbf{Fig. \ref{fig:sim1:A}}, the measure of adopters is the number of individuals with each belief in their knowledge graph, averaged over all beliefs, divided by the total population. Similarly, the measure of the susceptible population (and all discussion of the susceptible population) represents the fraction of individuals who would adopt each belief if exposed to it according to the appropriate decision rule for independent vs. interdependent diffusion, plus the fraction that has already adopted the belief.
\textbf{Fig. \ref{fig:sim1:B}} shows the Pearson correlation between the number of people who have adopted each belief at time $t$ and the number who were initially susceptible to the belief at $t=0$ but did not start with it. As this has no meaningful value at $t=0$, the curve is drawn from t1-t9.
\textbf{Fig. \ref{fig:sim1:C}} assesses the correlation between the number of individuals who have adopted each belief (knowledge graph edge) and the number who have adopted the most popular belief it shares a 'node' with, averaged over all beliefs.
\textbf{Fig. \ref{fig:sim1:D}} uses the clustering coefficient of a knowledge graph constructed from the most popular 10\% of beliefs as a demonstration that the most popular beliefs are mutually interrelated, and not merely all related to a single leading belief (\textit{e.g.} a star or barbell pattern). Clustering only makes sense when beliefs are conceptualized as some form of semantic network. Other conceptualizations of belief interaction might prefer to plot the number of top decile beliefs that each top decile belief interacts with. This measure gives essentially the same result (\textit{i.e.} large fractional growth over time in the interdependent case, with no change from randomness in the independent case) but fails to capture the mutual inter-relatedness indicated by the clustering coefficient. The measure is generally insensitive to the specific threshold used to define a ‘popular’ belief for any thresholds between about 5\% and 40\%. See the supplement for sensitivity analysis.

There are many complex measures of polarization in the literature \cite[for a sample]{baldassarri2007dynamics,dellaposta2015liberals,goldberg2018beyond,flache2011small,del2016echo,dandekar2013biased,vasconcelos2019consensus,spohr2017fake,tornberg2018echo,pennycook2019lazy}, which generally attempt to represent three basic intuitions. First, that individuals within the same ideological camp come to be more similar to one another. Secondly, that individuals in different ideological camps become more dissimilar to one another. Lastly, that an individual’s position on one dimension of belief becomes informative of their position on other dimensions. As my purpose is not to identify camps and their members, but to suggest that one set of conditions is more generative of polarization than another, these measures add more complexity than value. Instead I report heuristic measures characterizing the above three intuitions.

A simple and reproducible way to assess the similarity of individuals within an ideological camp – absent exogenous labels such as demographic or party – is to measure the similarity between all pairs of individuals and define a certain percentile as belonging to the same ideological camp. The more exclusive we are (\textit{i.e.} the higher the percentile), the more conservative the claim that these represent “within-camp” relationships. To define across-camp similarity, we can choose a percentile that (conservatively) represents relationships between individuals in different ideological camps. In \textbf{Figs. \ref{fig:sim2:A}, \ref{fig:experiment:B}, and \ref{fig:experiment:C}}, I use the $95^{th}$ and $5^{th}$ percentiles respectively. See the supplement for sensitivity analysis.

\textbf{Figs. \ref{fig:sim2:B}, \ref{fig:observational:C} and \ref{fig:experiment:A}} measure belief alignment as the percent of variation between individuals that can be explained by the best fitting axis in the space of possible beliefs, using singular-value decomposition. In \textbf{Fig. \ref{fig:sim2:B}}, the original feature space has one dimension for each belief in the simulation (300), and points representing each individual’s position in that feature space (60) according to the beliefs they have adopted. The offset graphs are exaggerated and show a larger population to illustrate how a component can explain more or less variation. In \textbf{Fig. \ref{fig:experiment:A}}, the feature space has 22 dimensions for the behavioral measures, and 11 dimensions for self-report measures (described below), and points representing 20 individuals.

\subsection*{Analysis of observational data}
The first column in \textbf{Fig. \ref{fig:observational}} shows simulated measures of belief clustering, inter-subject similarity, and percent variance in a given principal component over a range of thresholds and principal components, using the same simulation parameters described previously.

In order to assess clustering of popular beliefs in \textbf{Fig. \ref{fig:observational:A}}, a dataset needs to include a large diversity of connections between concepts. New York Times article data is taken from the KPTimes dataset\cite{gallina2019kptimes}, including 13 years of articles for all categories with $>= 3000$ documents. One line is shown for each of 27 news sections, each computed independently. When multiple categories were indicated, a document was assigned to the smallest category with $>= 3000$ documents. A weighted undirected graph was constructed for each category, in which article keywords formed nodes, and edges were weighted by the number of articles tagged with the two keywords in the edge. An alternative method in which edges are constructed by random draws without replacement gives the same qualitative effect, and is detailed in the supplement. Data on Web of Science paper keywords\cite{kowsari2017HDLTex} used the same analysis, showing one line for each of 7 major academic fields.

In order to assess similarity and political axis alignment in \textbf{Figs. \ref{fig:observational:B} and \ref{fig:observational:C}}, a dataset needs dense information about individual beliefs. Political opinion data comes from 58 political opinion questions from the Pew Political Typology Survey\cite{pew2014}, conducted in three waves in early 2014.
The social values analysis uses 144 questions regarding non-political social values in the World Values Survey\cite{wvsa2020} collected from 2017-2020. Missing, unknown, and refused responses ($<5\%$) are filled with the mean values for each question for each wave or country. As in the simulations above, similarity between individuals is assessed as the Pearson correlation of their responses to survey questions. Similarity curves are median-shifted in order to cleanly show the effect on all waves in the same plot. Principal components are computed from a space initially consisting of one dimension for each survey question. A full list of survey questions used is available in the supplement.

As we have no counterfactual world in which diffusion is assuredly independent, I compare the observed data to a synthetic control generated by shuffling beliefs among documents or individuals.

\subsection*{Experiment}
The preregistration for this experiment included all code necessary to conduct the experiment using Empirica\cite{almaatouq2021empirica} and perform all statistical analyses reported in this paper. The preregistration is available at \href{https://osf.io/239ns}{https://osf.io/239ns}.

Over the course of eight days in July 2020, I recruited 2768 U.S. and Canadian Mechanical Turk workers under the criteria that they were 18+ years old, and have completed 100+ HITs with a 90\%+ approval rating. Of these, 2400 completed training and were randomized into 20-person social networks. Each network was assigned to one of four (matched) experimental conditions, yielding $n=30$ samples per condition. This sample size was set by budget constraints. The participant population was 45\% female; mean 37.1 years old; 27\% high-school, 49\% bachelors, 16\% masters+ graduates. 96.8\% of players who completed training went on to complete all steps of the experiment, with less than 0.4\% difference in dropout between conditions.
The average payout was just over \$4, and the experiment took about 20 minutes.



Similarity between self-reported measures is assessed using Pearson’s correlation on the vectors of individuals’ responses. This measure has the advantage of being easily interpretable and having a well-defined range that is independent of the number of features in the vector of attributes being compared, and the negative region of which can be interpreted as expressing dissimilarity. To assess the similarity of the binary “behavioral” data I use the Phi coefficient, an analogous measure to Pearson’s correlation with the same interpretable range.

The behavioral measures in the experiment are sensitive not only to interdependence and network structure but also to the average level of diffusion of beliefs. To minimize noise due to differences in the level of activity between games, each of the behavioral measures is assessed compared to what would be expected due to chance, keeping the number of adopters of each clue and the number of clues adopted by each participant fixed. This correction was designed in simulation and preregistered.
}

\showmatmethods{} 

\acknow{Special thanks to Abdullah Almaatouq, Sinan Aral, Carolyn Fu, Hazhir Rahmandad, David Rand, Ray Reagans, Hagay Volvovsky, Duncan Watts, and the System Dynamics and Economic Sociology groups at MIT Sloan for their guidance. }

\showacknow{} 

\bibliography{references}

\end{document}


\maketitle
\tableofcontents

\section{Simulation}

\subsection{Simulation code}
The simulations presented in this paper can be replicated with the following code. This code is optimized for clarity over speed. This code was originally written using:
\begin{itemize}
  \item Python 3.7.1
  \item NetworkX 2.3
  \item Pandas 0.24.2
  \item Scikit-learn 0.20.1
\end{itemize}

In this code, \texttt{g} is the social network (a networkx graph object) with \texttt{n} agents. Each node in \texttt{g} is an agent numbered $0$\ldots$n-1$, and has an attribute \texttt{M} which stores the agent's current belief set as a knowledge graph (another networkx graph object for each agent). In the independent condition, agents also have an attribute \texttt{S} representing their susceptibility to beliefs, also formatted as a knowledge graph. When beliefs are passed between functions, it is either as an ordered tuple representing the edge in a knowledge graph \texttt{(2, 7)} or as a numpy array of tuples \texttt{np.array([(2, 7), (2, 16), \ldots])}

To import this module into another python file or jupyter notebook, call:
\begin{minted}{python}
    from example_code import *
\end{minted}
To run a matched simulation of interdependent and independent diffusion, call:
\begin{minted}{python}
    result = run()
\end{minted}
To run a number of simulations and average their output, call:
\begin{minted}{python}
    n_sims = 10
    df = pd.concat([run() for i in range(n_sims)])
    result = df.groupby(level=0).aggregate('mean')
\end{minted}
This code and supporting materials are available at: \url{https://github.com/JamesPHoughton/interdependent-diffusion}.

~
\hrule

\begin{minted}[linenos]{python}
"""
example_code.py
James Houghton
houghton@mit.edu
"""
import networkx as nx
import numpy as np
import itertools
import pandas as pd
import copy
from sklearn.decomposition import PCA


def susceptible(g, agent, belief):
    """Assess whether an agent is susceptible to a given belief"""
    if 'S' in g.node[agent]:  # has exogenous susceptibility defined (independent case)
        return g.node[agent]['S'].has_edge(*belief)
    else:  # interdependent case
        try:
            return nx.shortest_path_length(g.node[agent]['M'], *belief) <= 2
            # current holders are also susceptible
        except (nx.NetworkXNoPath, nx.NodeNotFound):
            return False  # no path exists between the nodes


def adopt(g, agent, belief):
    """Assess whether an agent will adopt a given belief"""
    suscep = susceptible(g, agent, belief)
    exposed = any([belief in g.node[nbr]['M'].edges() for nbr in g[agent]])
    return suscep and exposed  # both susceptibility and exposure required to adopt


def measure(g, beliefs, initial_susceptible=None, initial_adopted=None):
    """Take measurements of the state of the system (for creating figures)"""
    res = {}  # dictionary to collect measurements

    # Fig 2A: Susceptible and adopting populations
    # --------------------------------------------
    # build a matrix of who (rows) is susceptible to what beliefs (columns)
    suscep = pd.DataFrame(index=g.nodes(), columns=[tuple(b) for b in beliefs])
    for agent in g:
        for belief in suscep.columns:
            suscep.at[agent, belief] = susceptible(g, agent, belief)
    # return average susceptible fraction across all beliefs
    res['% susceptible'] = suscep.mean().mean()

    # build a matrix of who (rows) holds what beliefs (columns)
    adopt = pd.DataFrame(index=g.nodes(), columns=[tuple(b) for b in beliefs])
    for agent in g:
        for belief in adopt.columns:
            adopt.at[agent, belief] = g.node[agent]['M'].has_edge(*belief)
    # return average adopting fraction across all beliefs
    res['% adopted'] = adopt.mean().mean()

    # Fig 2B:correlation between predicted new adoption and actual new adoption
    # -------------------------------------------------------------------------
    if initial_adopted is not None and initial_susceptible is not None:  # t>0
        res['initial prediction correlation'] = np.corrcoef(
            adopt.sum(axis=0) - initial_adopted,
            initial_susceptible - initial_adopted
        )[1, 0]  # select an off-diagonal term
    else:  # first time => establish baseline
        initial_adopted = adopt.sum(axis=0)
        initial_susceptible = suscep.sum(axis=0)
        res['initial prediction correlation'] = np.nan  # measure has no meaning at t0

    # Fig 2C: correlation between a belief and it's most popular neighbor
    # -------------------------------------------------------------------
    adopt_counts = pd.DataFrame()
    adopt_counts['self'] = adopt.sum(axis=0)
    adopt_counts['leading neighbor'] = 0
    for c1 in adopt.columns:
        # search for the leading neighbor's popularity
        leading_value = 0
        for c2 in adopt.columns:
            if len((set(c1) | set(c2))) == 3:  # three nodes total => c1 and c2 are neighbors
                leading_value = max(leading_value, adopt_counts.loc[[c2], 'self'].values[0])
        adopt_counts.at[[c1], 'leading neighbor'] = leading_value
    res['leading neighbor correlation'] = adopt_counts.corr().loc['self', 'leading neighbor']

    # Fig 2D: clustering coefficient of 10% most popular beliefs
    # ----------------------------------------------------------
    # shuffle within sorted value so that when 10% falls within a level of popularity
    # we don't add spurious clustering by selecting sequential beliefs
    adopt_counts['shuffle'] = np.random.rand(len(adopt_counts))
    adopt_counts.sort_values(by=['self', 'shuffle'], inplace=True, ascending=False)
    leaders = adopt_counts.iloc[:int(len(adopt_counts) * 0.1)]  # take leading 10% of beliefs
    # construct knowledge graph from leading beliefs
    popular_graph = nx.from_edgelist(list(leaders.index))
    res['popular belief clustering'] = nx.average_clustering(popular_graph)

    # Fig 3A: similarity btw 5% and 95% most similar pairs
    # ----------------------------------------------------
    n_agents = len(adopt.index)
    trimask = np.tri(n_agents, n_agents, 0, dtype='bool')  # mask the diagonal and below
    corrs = adopt.astype(float).T.corr().mask(trimask).stack()
    res['95% similarity'] = np.percentile(corrs, 95)
    res['5% similarity'] = np.percentile(corrs, 5)

    # Fig 3B: PC1 percent variance
    # ----------------------------
    pca = PCA(n_components=1)
    pca.fit(adopt)
    res['PC1 percent of variance'] = pca.explained_variance_ratio_[0] * 100

    return res, initial_susceptible, initial_adopted


def simulate(g, n_steps=10):
    """Conduct a single run of the simulation with a given network"""
    # capture a list of all the beliefs in the population
    beliefs = np.unique([tuple(sorted(belief)) for agent in g
                         for belief in g.node[agent]['M'].edges()], axis=0)

    # measure initial conditions
    m0, initial_susceptible, initial_adopted = measure(g, beliefs)
    # initialize list to collect measurements at each time step
    m = [m0]

    # perform the simulation
    for step in range(n_steps):
        # cycle through agents in random order
        for ego in np.random.permutation(g):
            # cycle through all possible beliefs in random order
            for edge in np.random.permutation(beliefs):
                # check whether the selected agent adopts the selected belief
                if adopt(g, ego, edge):
                    # add the belief to the agent's knowledge graph
                    g.node[ego]['M'].add_edges_from([edge])
        # ignore returned init suscep and adopt
        m.append(measure(g, beliefs, initial_susceptible, initial_adopted)[0])

    return pd.DataFrame(m)  # format as pandas DataFrame


def run(n_agents=60, deg=3, n_concepts=25, n_beliefs=25, t_match_susceptibility=0, n_steps=10):
    """
    Run a matched pair of simulations (inter/independent) from the same initial condition

    Parameters
    ----------
    n_agents: (integer) - Number of agents in the population
    deg: (integer) - How many neighbors each agent has (on average)
    n_concepts: (integer) - How many nodes are in the complete semantic
                            network that beliefs are drawn from
    n_beliefs: (integer) - Exact number of beliefs (semantic net edges)
                           each agent is initialized with
    t_match_susceptibility: (integer) - time step at which the interdependent
                                        susceptibility will be matched
                                        (must be less than n_steps)
    n_steps: (integer) - Number of time steps in the simulation
    """

    # Shared Initial Setup
    # --------------------
    # create a random connected social network g0
    connected = False
    while not connected:
        g0 = nx.gnm_random_graph(n=n_agents, m=int(n_agents * deg / 2))
        connected = nx.is_connected(g0)

    # give agents their initial beliefs
    nx.set_node_attributes(
        g0,
        name='M',  # set node attribute 'M' (for 'mind')
        # create a knowledge graph with a different random set of beliefs
        # for each agent, and assign to nodes in the social network
        values={agent: nx.gnm_random_graph(n_concepts, n_beliefs) for agent in g0}
    )

    # Interdependent simulation
    # -------------------------
    g1 = copy.deepcopy(g0)  # create copy, to preserve initial conditions for other case
    res1 = simulate(g1, n_steps)

    # Independent simulation
    # ----------------------
    g2 = copy.deepcopy(g0)  # copy from original starting conditions

    # calculate the population likelihood of being susceptible to a given (non-held) belief
    p = ((res1.loc[t_match_susceptibility, '% susceptible'] - res1.loc[0, '% adopted']) /
         (1 - res1.loc[0, '% adopted']))

    # choose a set of beliefs for each agent to be susceptible to
    new_sus = {}
    for agent in g2:
        # potentially susceptible to any belief
        gc = nx.complete_graph(n_concepts)
        # temporarily remove existing beliefs
        gc.remove_edges_from(g2.node[agent]['M'].edges())
        # from remainder, randomly select a subset of beliefs to be susceptible to
        edges = list(itertools.compress(
            list(gc.edges()),  # selection candidates
            np.random.binomial(n=1, p=p, size=len(gc.edges())) == 1  # selection mask
        ))
        # add susceptibility to existing beliefs
        edges += list(g2.node[agent]['M'].edges())
        # create networkx graph of susceptibilities
        new_sus[agent] = nx.from_edgelist(edges)

    # assign new susceptibilities to positions in social network
    nx.set_node_attributes(g2, name='S', values=new_sus)

    # perform independent simulation
    res2 = simulate(g2, n_steps)

    return pd.merge(res1, res2, left_index=True, right_index=True,
                    suffixes=(' (inter)', ' (indep)'))  # format as single DataFrame

\end{minted}

\hrule

\subsection{Motivating the ``Knowledge Graph'' representation of belief interaction}
There are two primary ways that we could model the interaction between beliefs. One method used by Friedkin et al. \cite{friedkin2016network} is to create a matrix of the compatibility between candidate beliefs (rows) and preexisting beliefs (columns). In this model, weights in the matrix determine the influence of (unstructured) existing beliefs on the probability of an individual adopting each candidate belief. This formulation assumes external `logic' under which certain beliefs naturally go together.

A second method is the knowledge graph representation suggested by Goldberg and Stein \cite{goldberg2018beyond} and by Schilling \cite{schilling2005small}, in which existing beliefs have no independent effect on an an agent's susceptibility to a candidate belief. Instead, they influence future adoption decisions through the structure they create in the individual's knowledge graph. In this representation, all beliefs are potentially compatible with one another, depending on the arrangement of other beliefs that the individual holds.

When we study the effect of diffusion on the emergence of worldviews and polarization, it's important that we choose a representation of belief interaction that does not foreordain particular outcomes. If we can predict patterns of belief clustering and polarization from the decision logic alone, then the simulation will be unable to identify whether clustering is shaped by social contagion, or is merely a result of the assumed decision logic.

Unfortunately, an interdependence matrix which maps the presence of belief A to the likelihood of adopting belief B will always be informative of the final configuration of beliefs (or trivial). We can demonstrate this outcome with an extremely simple deterministic model with no diffusion at all. This model is in the style of Friedkin et al.\cite{friedkin2016network}, but omits social influence, stochasticity, and degrees of belief. These simplifications let us see intuitively why the interdependence matrix is problematic for our purpose. When stochasticity, social influence, etc. are present, the problem doesn't go away, it is just masked by the complexity of the model. In this simplified interdependence-matrix belief adoption model:

\begin{enumerate}
    \item Each agent is exposed to all beliefs in every timestep (making this an individual learning model, not a social learning model).
    \item The presence of belief ‘A’ in an agent's existing belief set either contributes (+) to an agent's likelihood of adopting candidate belief ‘B’, or takes away from it (–).
    \item The absence of belief ‘A’ from an agent's belief set has no direct influence on an agent's likelihood of adopting candidate belief ‘B’.
    \item Belief adoption is binary and permanent.
    \item Agents adopt a candidate belief if the majority of their current beliefs support doing so.\footnote{Other thresholds would lead us to the same insights that we will develop with the ``majority rule''.}
\end{enumerate}

\begin{figure}[h!]
\centering
\includegraphics[width=1.0\columnwidth]{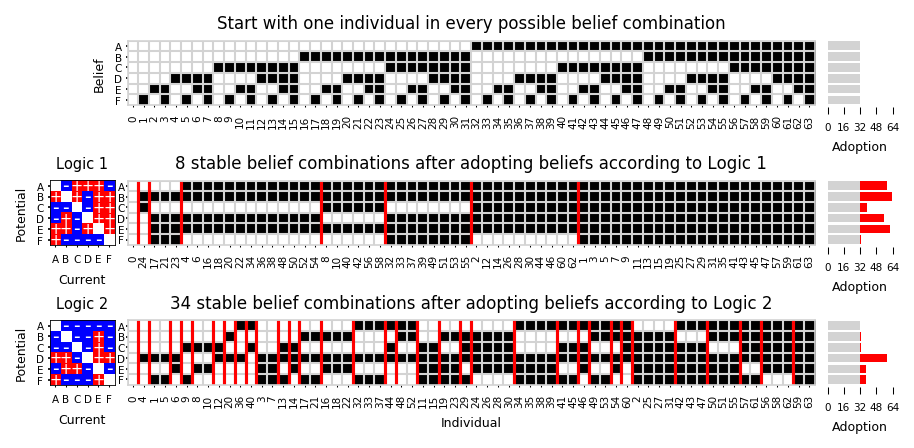}
\caption{With an interdependence-matrix style belief structure, polarization and belief clustering can be explained by the assumed compatibility relationships. But where does this logic come from?}
\label{fig:rule_matrix}
\end{figure}

To show that the matrix of influence is informative of the outcomes of contagion, we simulate a population of 64 individuals, each with a unique combination of six beliefs, denoted A-F. The top center chart in Fig. \ref{fig:rule_matrix} illustrates this starting condition. Each column represents an individual agent (0-63), each row represents a belief (A-F). I darken the corresponding square to show that an agent has adopted a particular belief. To the right-hand side of the adoption plot, a histogram shows the total number of individuals adopting each belief. In the initial condition, all beliefs have been adopted by 32 individuals.

The subsequent two plots in Fig. \ref{fig:rule_matrix} show the final adoption of beliefs under two influence matrices (Logic 1 and Logic 2) in the style of Friedkin et al.\cite{friedkin2016network}. Logic 1 and Logic 2  are randomly generated influences between beliefs, shown as + (red) and – (blue). In each matrix, if an individual holds A, then the influence of that belief on the adoption of belief B is found in Column A, Row B. In Logic 1, the influence of A on B is positive, and in Logic 2, this influence is negative.

For each individual, we calculate the net influence of existing beliefs on adoption of each candidate belief. If the candidate belief gets a majority of + votes then it is adopted. We repeat the process until all agents have converged on a stable combination of beliefs.

Any model of social learning with binary and permanent belief adoption and a finite number of beliefs will show some form of belief consolidation, and an increase in mean similarity between individuals. We expect to see that from our maximally-differentiated initial conditions, some groups of stable belief combinations should emerge. In the adoption plots in Fig. \ref{fig:rule_matrix}, I group individuals according to their stable sets of beliefs, indicating the groups with a red divider.

Despite being drawn from the same pool of possible logics, the two Friedkin-style logics yield very different stable combinations of beliefs. In Logic 1, there are 8 different stable combinations, and in Logic 2, there are 34. In a diffusion model, we would expect to see a lot more consolidation of beliefs, and clustering of individuals, using Logic 1 rather than Logic 2, regardless of social contagion. Moreover, the two different logics strongly influence which beliefs we should expect to be widely adopted in the population. Logic 1 suggests strong new adoption (shown in red) of beliefs A, B, D, and E. Logic 2 shows that only belief D is widely adopted, with belief A having no new diffusion at all.

In the real world, it may well be that a natural logic ties some beliefs together, and promotes the diffusion of some beliefs at the expense of others. In our simulation, however, this influence makes it difficult to identify clustering and amplification of beliefs that is due to social contagion. In this paper, I have suggested an alternative formulation, in which beliefs are formalized as edges in a knowledge graph, and the adoption rule is that beliefs are adopted if they close a triangle in an individual’s knowledge graph.

\begin{figure}[h!]
\centering
\includegraphics[width=1.0\columnwidth]{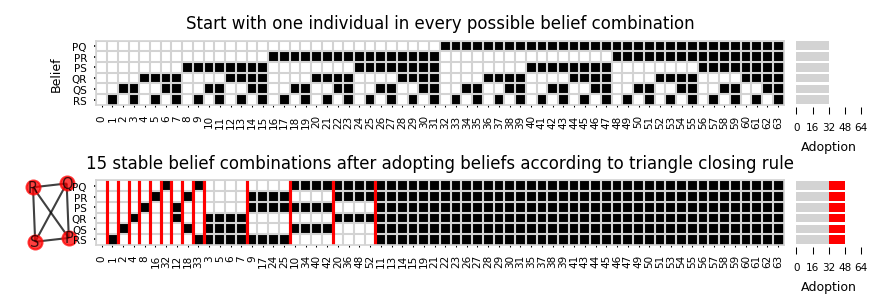}
\caption{With the knowledge graph representation of belief structure, we have confidence that any variance in adoption patterns, or differences in social clustering, are due to the interaction of the belief structure with the diffusion process.}
\label{fig:rule_triangle}
\end{figure}

In Fig \ref{fig:rule_triangle}, I repeat the above analysis with this new formulation. Again there are six beliefs, representing all of the possible edges in a knowledge graph with concepts P, Q, R and S. Again, each of 64 individuals is initialized with a unique combination of beliefs, and has access to all six beliefs. Each individual adopts new beliefs that are consistent with the triangle-closing decision rule, and I plot the stable belief combinations after each individual has individually converged. As there is only one decision rule, there can be only one outcome for grouping the sets of stable beliefs.

With this formalization, we know exactly how the decision rule for adoption influences the number of possible stable states; and most importantly, these states are symmetric with respect to the individual beliefs. The histogram to the right of the second row of \ref{fig:rule_triangle} shows that each belief has an equal number of new adoptions, and that we should not expect any beliefs to preferentially diffuse as a result of the decision rule itself.

In this paper I have shown that some beliefs do indeed diffuse much more widely than others, and that the population clusters into a subset of the possible stable states. Because the decision rule does not influence which of the beliefs diffuse most widely, or suggest variation in the number of social clusters we can expect, we have confidence that the results are genuinely due to the interactions between beliefs as they diffuse.

\subsection{Sensitivity to parameters}
\label{section:thresholds}
\begin{figure}[h!]
\centering
\includegraphics[width=0.5\columnwidth]{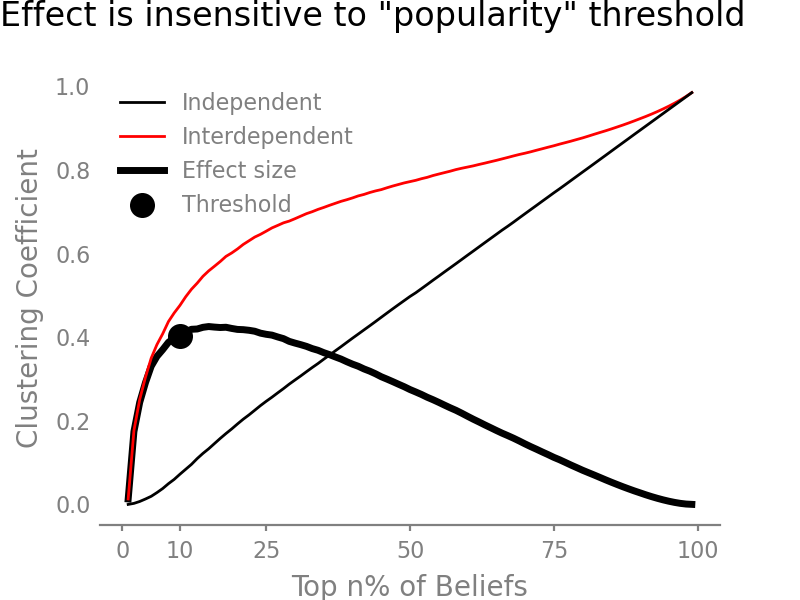}
\caption{The effect of interdependence on the clustering of the most popular beliefs is robust to a wide range of thresholds for ``popularity''. In the simulations presented in this experiment, I use a conservative 10\% threshold.}
\label{fig:clustering_threshold}
\end{figure}

Figure 2D in the main body of the paper makes a comparison between the clustering coefficient amongst the most popular beliefs, defined as the 10\% most widely adopted. Fig. \ref{fig:clustering_threshold} in this supplement shows that the qualitative result is insensitive to this particular choice of threshold between approximately 5\% and 40\%.

\begin{figure}[h!]
\centering
\includegraphics[width=0.6\columnwidth]{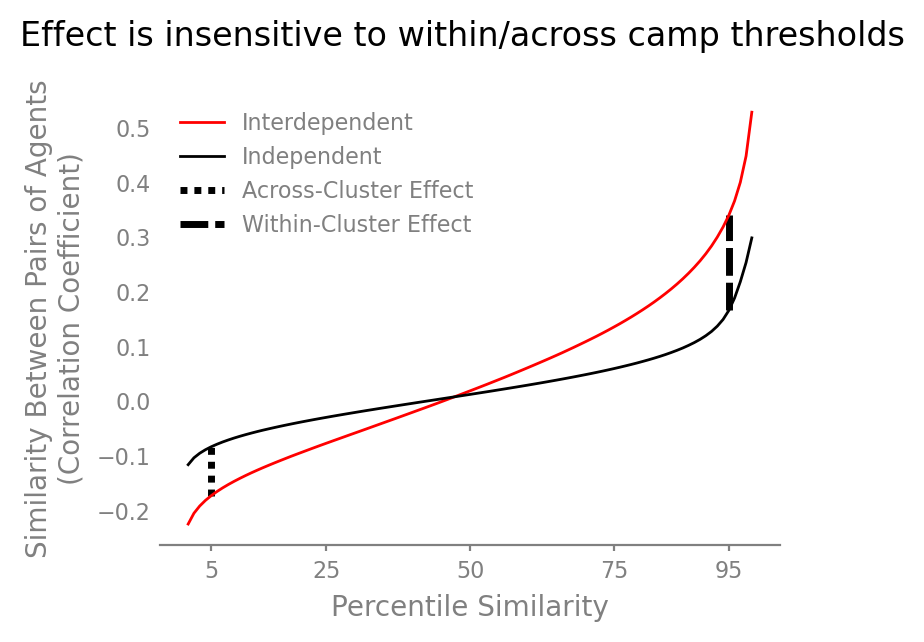}
\caption{The effect of interdependence on within-camp and across-camp similarity is robust to a wide range of thresholds for assessing whether relationships are within or across camp. In the simulation and experiment, I use conservative 5\textsuperscript{th} and 95\textsuperscript{th} percentile thresholds.}
\label{fig:camp_threshold}
\end{figure}

In Fig. 3A, 5B and 5C of the paper, I use 5\textsuperscript{th} and 95\textsuperscript{th} percentile values to characterize the level of similarity across and within ideological camps. These thresholds are arbitrary, in that we could have plausibly used a wide range of other values. Fig. \ref{fig:camp_threshold} shows that the qualitative effect we are interested in is insensitive to the precise choice of threshold.

\section{Analysis of Observational Data}

\subsection{Description of Datasets}
All code needed to generate Fig. 4 in the main body is available in the \href{https://github.com/JamesPHoughton/interdependent-diffusion/tree/master/observational}{/observational} folder of the code repository.

\subsubsection{KPTimes}
The KPTimes dataset \cite{gallina-etal-2019-kptimes} contains New York Times articles between January 1, 2006 and June 25, 2019. The analysis in Fig. 4 of the main body included 288,545 documents from 27 categories. These categories were selected on the criteria that they include $>= 3000$ documents. When a document was labeled with multiple categories, it was assigned to the smallest category with $>= 3000$ documents. The categories included in the analysis are: (parentheses indicate the number of documents per category)

\noindent
\texttt{africa (5203), americas (4801), asia (16218), baseball (12754), basketball (7023), business (33686), dealbook (3815), economy (3186), europe (14630), football (8209), global (5762), golf (3471), hockey (4323), media (9427), middleeast (12338), national (3513), ncaabasketball (4011), ncaafootball (3180), nyregion (43221), politics (15094), science (7419), soccer (5254), sports (15059), technology (9634), tennis (3520), us (29420), world (4374)}

\subsubsection{Web of Science}
The Web of Science dataset \cite{WOSv6} was originally collected by Kowsari et al. \cite{kowsari2017HDLTex} for training text classification algorithms, and was repurposed for this analysis. The datset contains 46,985 documents in 7 Major fields: (parentheses indicate the number of documents per category)

\noindent
\texttt{Computer Science (6514), Civil Engineering (4237), Electrical Engineering (5483), \\Mechanical Engineering (3297), Medical (14625), Psychology (7142), Biochemistry (5687)}

\subsubsection{Pew Political Typology Survey}
The Pew Political Typology Survey interviewed 10,013 US adults in three approximately equal waves from January 23, 2014 through March 16, 2014. Each wave was split into half land-line and half cellphone interviews. The analysis in Fig. 4 of the main body included 58 questions about political opinions, chosen prior to data observation to focus only on political opinions (i.e. to exclude questions about demographics and other ascribed characteristics) and to consolidate redundant questions. Not all questions were included in each wave. The question codes included are:

\noindent
\texttt{q25a, q25b, q25c, q25d, q25f, q25g, q25h, q25i, q25j, q25k, q25l, q25m, q25n, q25o, q25p, q50q, q50r, q50s, q50t, q50u, q50v, q50w, q50y, q50z, q50aa, q50bb, q50dd, q50ee, q50ff, q50gg, q50hh, q51ii, q51jj, q51kk, q51ll, q51mm, q51oo, q51pp, q53, qb26, qc26, qb106, qb107, qb109, qb110, qc115, qc116, q121, q122, q123, q124, q125, q126, qc127, qc128, qc135, qc56, qb108}

\subsubsection{World Values Survey}
The 2020 World Values Survey \cite{wvs2020} contains 70,867 responses collected from 2017-2020. The analysis in Fig. 4 of the main body used 144 questions relating to apolitical social values. The questions to be used were chosen prior to data observation to focus only on social opinions (i.e. to exclude questions about demographics and other ascribed characteristics, and to exclude explicitly political questions) and to consolidate redundant questions. The question codes included are:

\noindent
\texttt{Q1, Q2, Q3, Q4, Q5, Q6, Q7, Q8, Q9, Q10, Q11, Q12, Q13, Q14, Q15, Q16, Q17, Q18, Q19, Q20, Q21, Q22, Q23, Q24, Q25, Q26, Q27, Q28, Q29, Q30, Q31, Q32, Q33, Q34, Q35, Q36, Q37, Q38, Q39, Q40, Q41, Q43, Q44, Q58, Q59, Q60, Q61, Q62, Q63, Q64, Q65, Q66, Q67, Q68, Q69, Q70, Q71, Q72, Q73, Q74, Q75, Q76, Q77, Q78, Q79, Q80, Q81, Q82, Q83, Q84, Q85, Q86, Q87, Q88, Q89, Q106, Q107, Q108, Q109, Q110, Q121, Q122, Q123, Q124, Q125, Q126, Q127, Q128, Q129, Q130, Q158, Q159, Q160, Q161, Q162, Q163, Q164, Q165, Q166, Q167, Q168, Q169, Q170, Q171, Q172, Q173, Q174, Q176, Q177, Q178, Q179, Q180, Q181, Q182, Q183, Q184, Q185, Q186, Q187, Q188, Q189, Q190, Q191, Q192, Q193, Q194, Q195, Q196, Q197, Q198, Q235, Q236, Q237, Q238, Q239, Q241, Q242, Q243, Q244, Q245, Q246, Q247, Q248, Q249}

The analysis in Fig. 4 of the main body includes a curve for each of 49 different country codes: (parentheses show number of respondents)

\noindent
\texttt{AND (1004), ARG (1003), AUS (1813), BGD (1200), BOL (2067), BRA (1762), CHL (1000), CHN (3036), COL (1520), CYP (1000), DEU (1528), ECU (1200), EGY (1200), ETH (1230), GRC (1200), GTM (1203), HKG (2075), IDN (3200), IRN (1499), IRQ (1200), JOR (1203), JPN (1353), KAZ (1276), KGZ (1200), KOR (1245), LBN (1200), MAC (1023), MEX (1739), MMR (1200), MYS (1313), NGA (1237), NIC (1200), NZL (1057), PAK (1995), PER (1400), PHL (1200), PRI (1127), ROU (1257), RUS (1810), SRB (1046), THA (1500), TJK (1200), TUN (1208), TUR (2415), TWN (1223), UKR (1289), USA (2596), VNM (1200), ZWE (1215)}

\subsection{Alternate clustering analysis}
The analysis presented in Fig. 4A of the main body included connections between all keywords attached to each newspaper article or academic paper. An alternate method of producing these graphs is to only include connections between \textit{pairs} of keywords within each article, randomly selected without replacement. While this results in a significantly sparser dataset, it guarantees that none of the clustering observed is due to clustering within documents.

\begin{figure}[H]
\centering
\includegraphics[width=0.8\columnwidth]{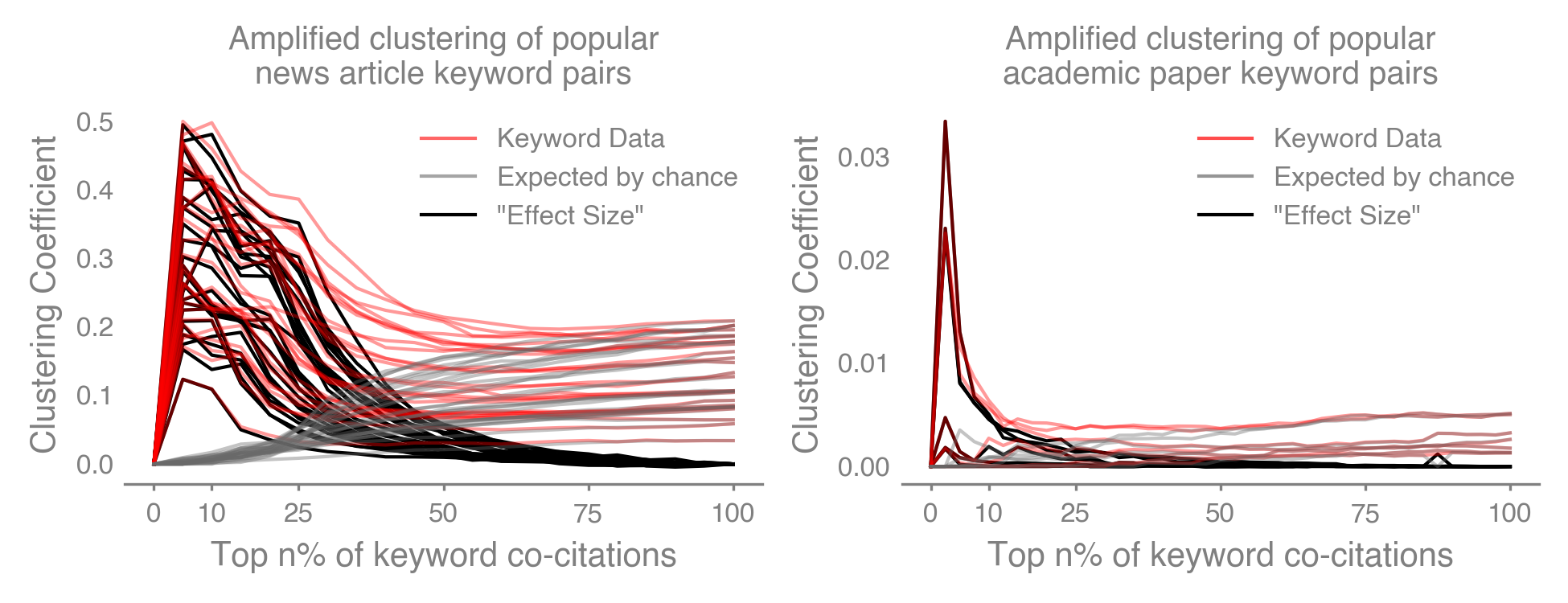}
\caption{The predicted amplification of clustering among popular keyword pairs is preserved when the data is restricted to disallow any contribution to clustering from the attachment of more than two keywords to the same article.}
\label{fig:pairwise_clustering}
\end{figure}

With this method, the overall clustering is reduced across all thresholds, and in particular thresholds that include keywords found in only a few documents. However, the qualitative prediction of increased clustering among the most popular connections between keywords remains, as shown in Fig. \ref{fig:pairwise_clustering}.

\section{Experiment}
The experimental design had two major objectives. 1) Allow for the manipulation of interdependence between diffusants, and 2) minimize the spillover of real-world polarization into the artificial social networks of the experiment. Beyond those considerations, I wanted to use a set of diffusants that individuals could quickly understand and would be interested in. A detective case solving a burglary was sufficiently relatable and abstract from the real world to serve as a stimulus and manipulation. Section \ref{design_considerations} details the various experiment design choices that were not specific to the manipulation. Section \ref{interface} details the participant's experience, and Section \ref{manipulation} details the design of the mystery's clues, which formed the experimental manipulation. Section \ref{data_collection} describes how data was collected and handled. Section \ref{prereg} describes the experiment's preregistration and deviations between the preregistration and the realized experiment. Section \ref{outcomes} lists outcomes of the experiment that have not already been discussed.

\subsection{Design considerations and experiment parameters}
\label{design_considerations}
There are many ways that the game interface and experimental parameters could have been implemented. Each decision took into account the feasibility of implementation, the constraints of the online lab context, the desire to create a naturalistic information and task environment, and the desire to cleanly isolate the mechanisms being tested. Below I justify the major decision points.

\textbf{Information was presented to each participant all at once} in the form of Detectives' Notebooks shared by their neighbors. This was an intuitive structure for the detective-game task; it allowed players to quickly understand how their categorization decisions would be shared and how the information was presented to them. There is a slight cost in real-world representativeness with this design choice, as it is different from how information is typically presented in social media. However, the alternative ``scrolling feed'' type information display has recency and primacy effects, and opens questions about how we should aggregate social information from multiple players. Showing all information at once, in the same order that it is sorted by the neighbor, eliminates the effect of alternate ordering sequences.

\textbf{Individuals had three social network neighbors.}
Given the chosen display, the number of neighbors is limited by the size of the screen and an individual’s ability to process information. The minimum number of neighbors for a non-trivial social network is 3, and is also a reasonable number for managing the cognitive load in the game.

\textbf{Individuals began with four clues in their detectives' notebooks.}
Fewer starting clues are preferred for minimizing individuals' cognitive load. With three neighbors, individuals began the game having to process 16 clues. The next increment (5 starting clues each) would have given 20 items for an individual to process at game start, which (in ``friends and family'' beta tests) proved to be cognitively overwhelming.

\textbf{The social network contained 20 players.}
Larger numbers of players are better for generalizability and seeing an effect size. On the other hand, smaller networks allow more replications and are easier to recruit and coordinate. There needed to be enough players that the mean shortest-path-length was greater than two, to realistically represent multi-stage diffusion. Pilot tests showed that we could reasonably expect to fill four 20-player games at once, and simulation suggested that 20-player networks should be sufficient to detect an effect size.

\begin{figure}[h!]
\centering
\includegraphics[width=0.9\columnwidth]{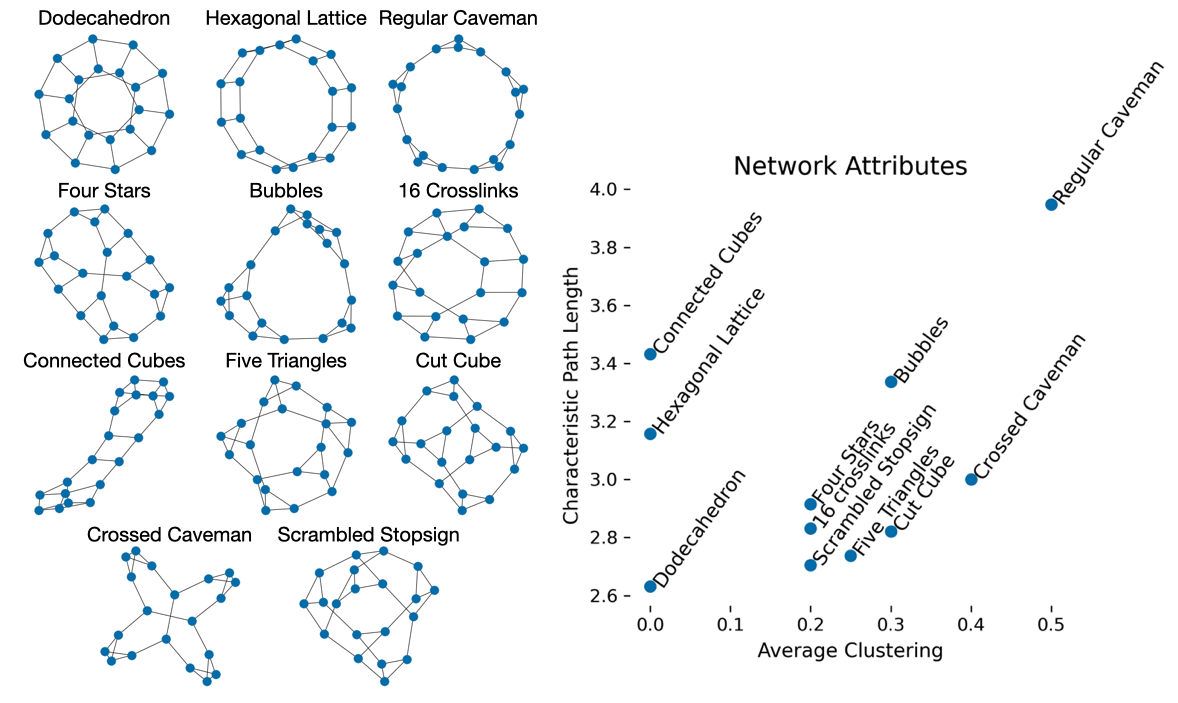}
\caption{Various possible social networks were evaluated, and two were selected (dodecahedron and regular caveman) to maximize the difference in average clustering and characteristic path length.}
\label{fig:all_networks}
\end{figure}

\textbf{The social networks were shaped as a dodecahedron and a regular connected caveman (k=5) network.}
I evaluated eleven symmetric candidate social networks (n=20, degree=3) shown in Fig. \ref{fig:all_networks}. Of this set, the dodecahedral network minimizes the average shortest path between individuals with no network clustering, and represents a social network in which \textit{a priori} we should expect to see low polarization. A regular connected caveman network maximizes the characteristic path length and exhibits strong clustering, and so we expect to see more polarization in this network. Edgelists for each of the eleven social networks are are available in the \href{https://github.com/JamesPHoughton/interdependent-diffusion/tree/master/experiment/setup}{/experiment/setup} folder of the code repository.

\textbf{Each game contained 78 unique clues.}
From an information diversity perspective, more clues is better. With 4 starting clues and 20 players, we can have up to 80 unique clues in the game. A complete 13 node clue network (knowledge graph) has 78 clues, and the two spots remaining were filled with the (given) link between the crime scene and the stolen object.

\textbf{Each clue was represented in exactly 1 starting notebook.}
In order to make sure that the initial frequency of information in the network did not bias the network to certain outcomes, each clue was present exactly once in the clues initially assigned to players. Other than this constraint, clues were assigned randomly. The clue linking the stolen object to the crime scene was included 2 additional times to fill out the 80 slots (20 players * 4 starting clues) available.

\begin{figure}[h!]
\centering
\includegraphics[width=.5\columnwidth]{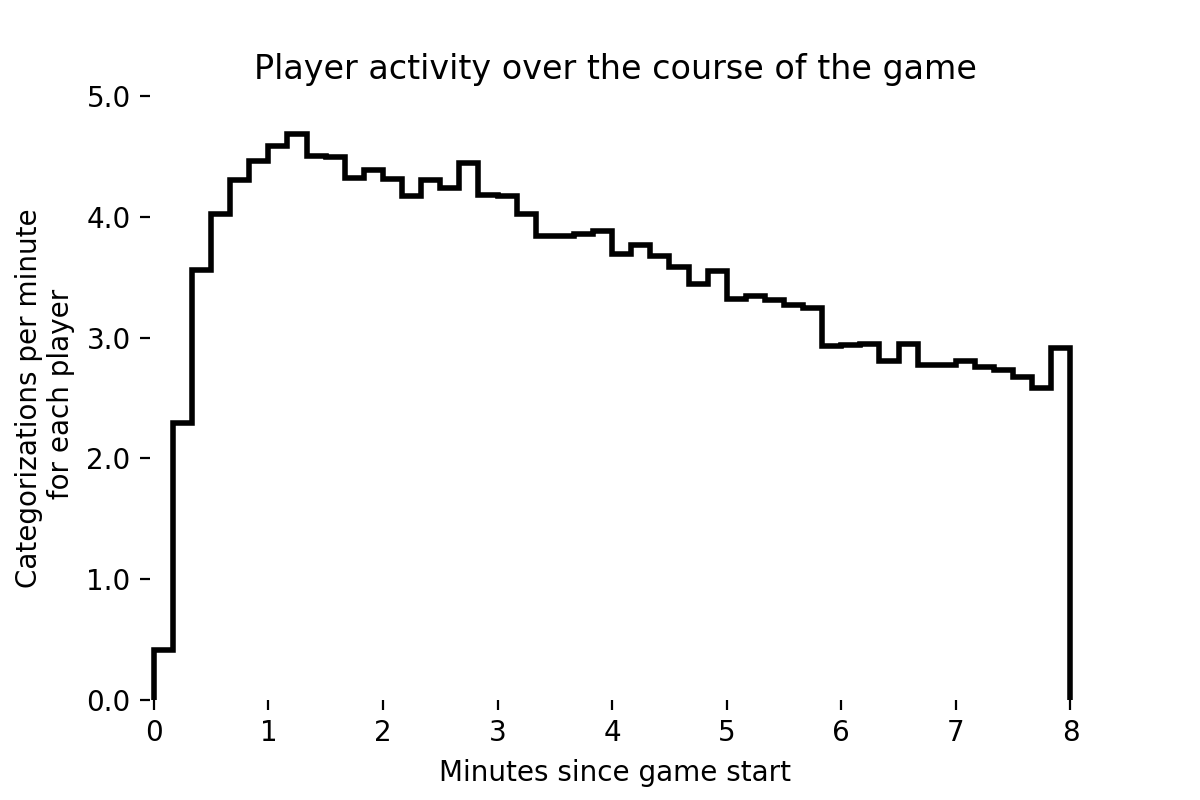}
\caption{Average rate of activity for all players in experiment runs (excluding pilots).}
\label{fig:player_activity}
\end{figure}

\textbf{The game was played for 8 minutes.}
Games needed to be long enough that participants had a chance to sort clues and make sense of the mystery, but short enough that they remained engaged and did not drop out of the experiment. Varying the length of games during pilot trials suggested that 8 minutes was a good balance between these constraints. Fig. \ref{fig:player_activity} shows that the average rate of categorization activity built over the first minute as players started to develop an understanding of the mystery, and then declined as individuals settled in on a solution.

\textbf{Post-game opinions were elicited as likelihood of participation for each element in the mystery.}
Rather than force individuals to make a discrete choice between suspects/vehicles etc., individuals assessed how likely each element of the mystery was to have been involved in the crime. While this is different from how a jury might assess the guilt of a suspect in a courtroom, it gave more resolution on the strength of individuals' opinions.

\textbf{Each condition was sampled 30 times.}
Each sample of the four conditions in this experiment cost about \$400. As there is no precedent for this type of multi-player, multi-diffusant experiment, and I did not have enough pilot data to know how effect sizes predicted in simulation would translate to experiment, it was difficult to assess the marginal benefit of additional samples relative to their cost. Based on the technical success of the pilots, I was able to secure \$12,000 to compensate participants, allowing for 30 sets of data-points.

\subsection{Design of participant interface}
\label{interface}
The experiment was implemented using the Empirica \cite{almaatouq2021empirica} framework, a platform for realtime multiparticipant web-based experiments.

\subsubsection{Consent}
\begin{figure}[h!]
\centering
\fbox{\includegraphics[width=0.6\columnwidth]{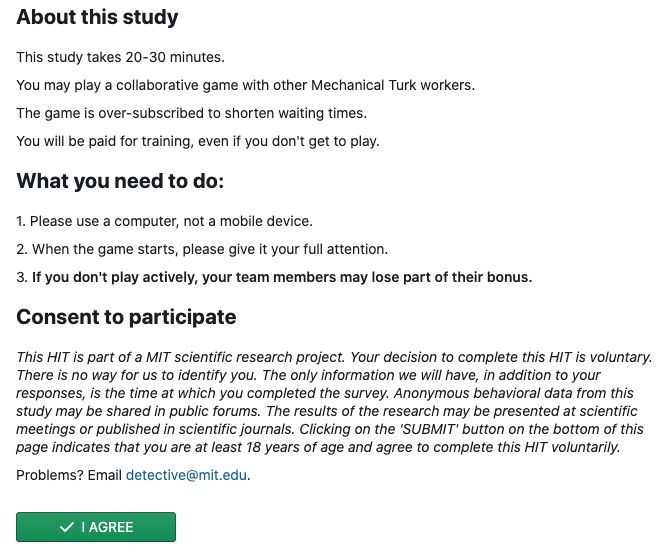}}
\caption{Participants indicate consent before proceeding.}
\label{fig:consent}
\end{figure}

Upon arrival, participants were shown a consent screen (Fig. \ref{fig:consent}) telling them about the study and what they would be expected to do, including information that the games were oversubscribed and that they may not be able to complete the whole game. Those who gave their consent to participate continued to training.

\subsubsection{Training}

\begin{figure}[h!]
\centering
\fbox{\includegraphics[width=.7\columnwidth]{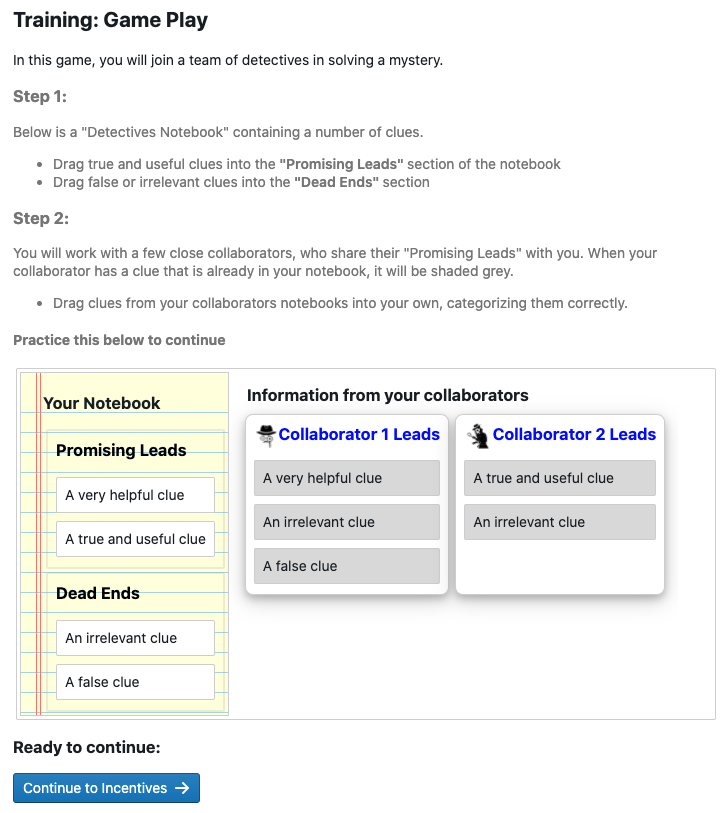}}
\caption{Training screen 1: How to use the game interface}
\label{fig:training1}
\end{figure}

The first training screen (Fig. \ref{fig:training1}) instructed participants in how to interact with the Detective Game interface. They were asked to sort clues into ``Promising Leads'' and ``Dead Ends'' by dragging them into labeled sections of their ``Detective’s Notebook''. Having done so, they were shown clues that two artificial ``collaborators'' had categorized as promising leads. Each participant then had to correctly sort the collaborators' clues before they could continue to the next training screen.

\begin{figure}[h!]
\centering
\fbox{\includegraphics[width=.7\columnwidth]{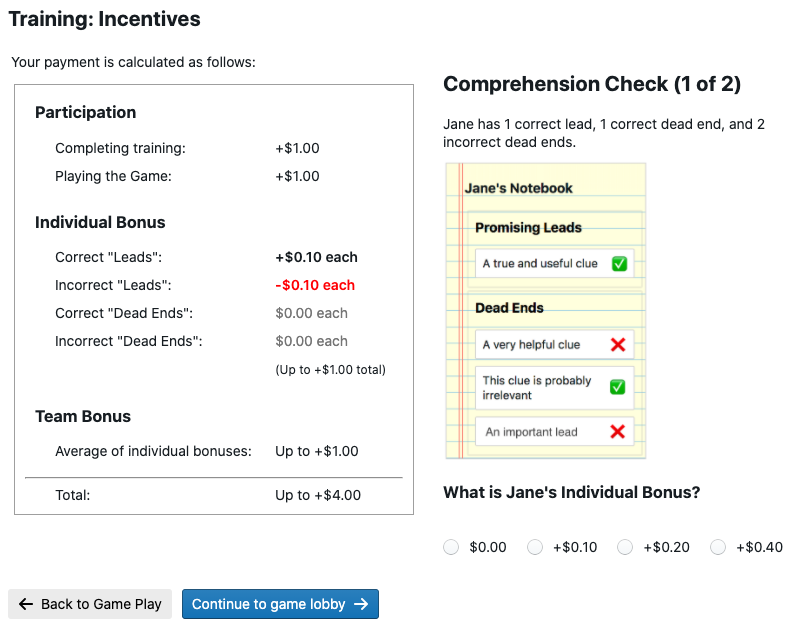}}
\caption{Training screen 2: Incentives and comprehension check}
\label{fig:training2}
\end{figure}

The second training screen (Fig. \ref{fig:training2}) told participants how they would be rewarded for their performance. Individuals were told that they would receive \$0.10 for each clue correctly categorized as a promising lead, and would be penalized \$0.10 for each clue categorized as a promising lead that was actually false. They were also told that they would be rewarded for their team’s average performance, receiving the average of all players’ individual bonuses as a Team Bonus. These incentives encourage individuals to carefully sort each clue according to their best estimate of its veracity, and to share clues with their neighbors that they believe will improve the team’s collective sense-making ability. Setting the reward for success to be equal to the penalty for mistakes encourages participants to accurately assess each statement, rather than ‘hedge’ by keeping too many or too few clues. Participants completed a comprehension check to ensure that they understood their incentives before they could proceed to the game. Participants were compensated \$1 for training.

\subsubsection{Game introduction}
After completing training (taking between 2 and 4 minutes), participants entered a waiting room until enough players had completed training and were ready to play.\footnote{Participants in each block were randomized into two sets of 40, each of which was then split again upon launch. This was due to the history of how the game was developed, and how the implementation of the game and the conditions of the experiment evolved over time. In future implementations, it would be simpler to have one condition per group, although the results would be identical.} The training was oversubscribed so that if some participants were unable to complete the training the game could still launch.

\begin{figure}[h!]
\centering
\fbox{\includegraphics[width=0.7\columnwidth]{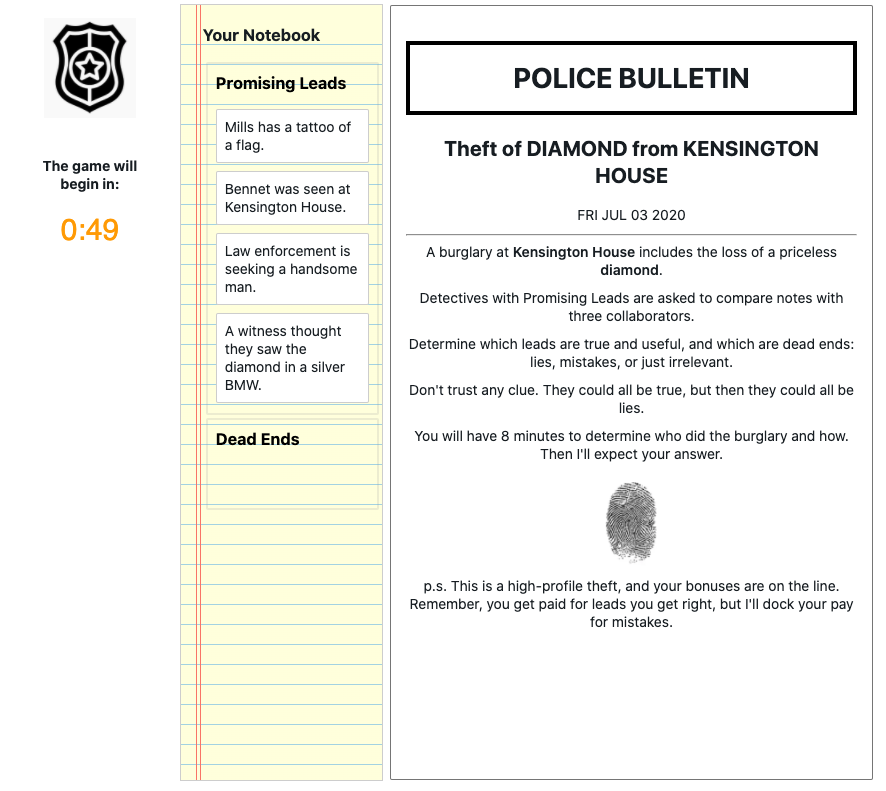}}
\caption{Exposition: Players were introduced to the case and reminded of their task when the game launched.}
\label{fig:exposition}
\end{figure}

When the game filled, players were assigned to locations in their social network. Each individual was given a Detective’s Notebook with four randomly assigned clues in the Promising Leads section. They were shown a “Police Bulletin” (Fig. \ref{fig:exposition}) which gave them background information about the mystery and reminded them of their task. They had 60 seconds to view this information and orient to the task.

\subsubsection{Playing the game: Exchanging clues}
\begin{figure}[h!]
\centering
\fbox{\includegraphics[width=1.0\columnwidth]{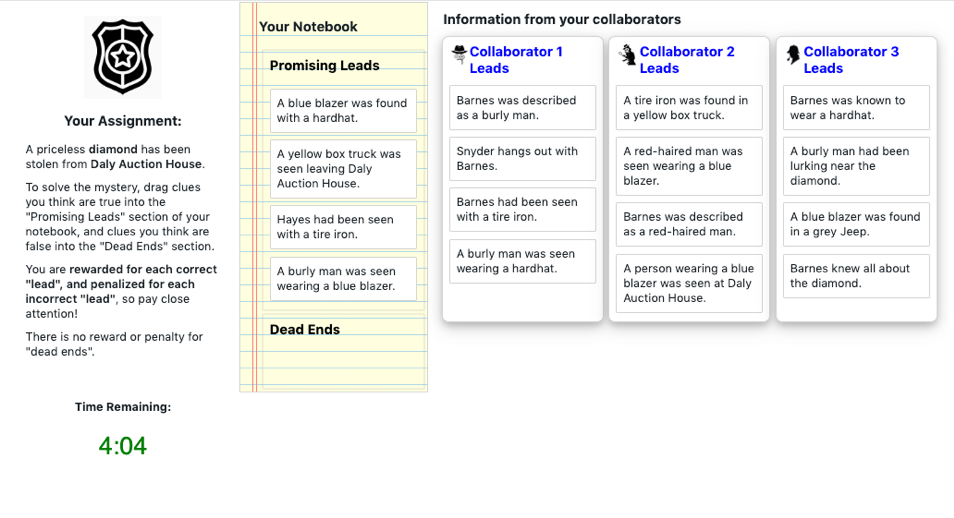}}
\caption{Player interface with interdependent clue set}
\label{fig:interdependent_game}
\end{figure}

\begin{figure}[h!]
\centering
\fbox{\includegraphics[width=1.0\columnwidth]{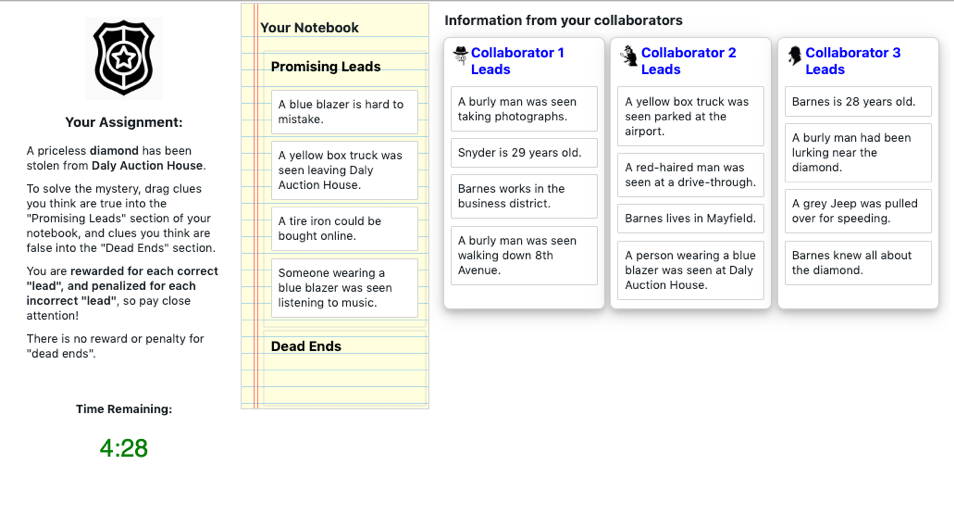}}
\caption{Player interface with matching independent clue set}
\label{fig:independent_game}
\end{figure}

When the game launched, the police bulletin was replaced with the promising leads sections of three collaborators’ notebooks, showing the participants a total of 16 unique clues at the start of the game. Individuals at corresponding positions in each social network were given clues that were as similar as possible while allowing for the intervention. These are shown for players in the treatment and control conditions in Figs. \ref{fig:interdependent_game} and \ref{fig:independent_game} respectively.

The game was played in real-time over 8 minutes. When a participant made any change to their promising leads, their neighbors immediately saw the change on their own screen. The starting clues of every individual were recorded, and every change to every player’s detective notebook was logged, such that the state of every player’s notebook can be reconstructed at each moment in the game. Participants were compensated \$1.00 for playing the game.

\subsubsection{Post-game survey: Making the case}

\begin{figure}[h!]
\centering
\fbox{\includegraphics[width=0.8\columnwidth]{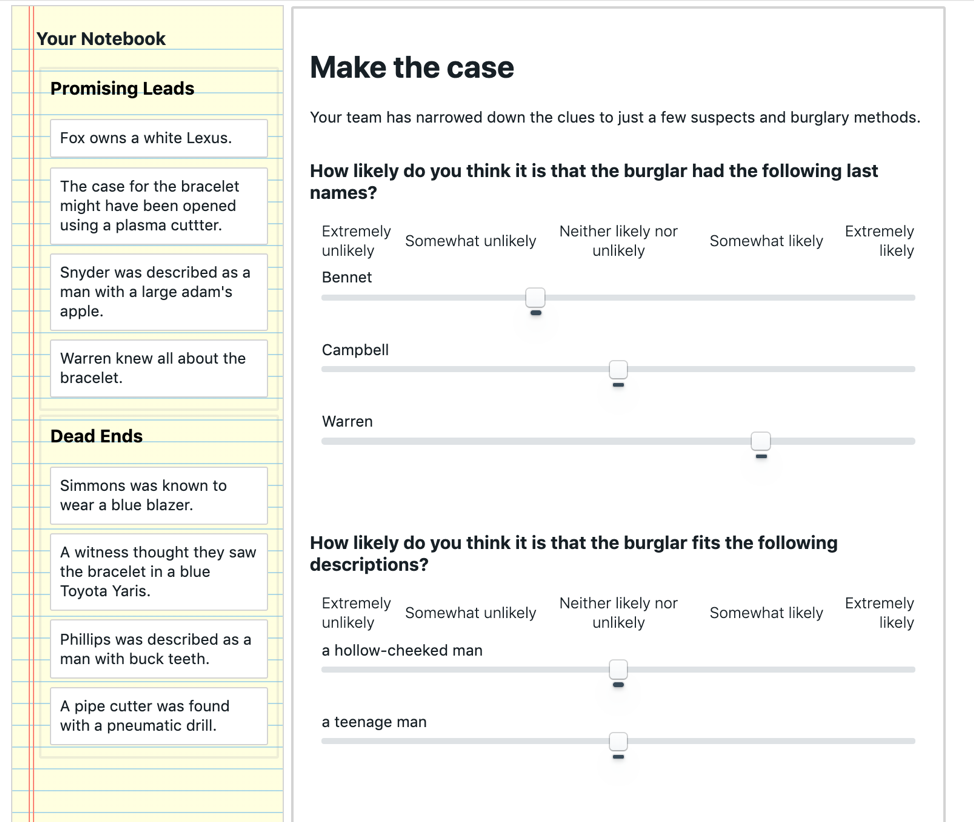}}
\caption{Post-game survey: Make the case for who committed the burglary}
\label{fig:make_the_case}
\end{figure}

\begin{figure}[h!]
\centering
\fbox{\includegraphics[width=0.8\columnwidth]{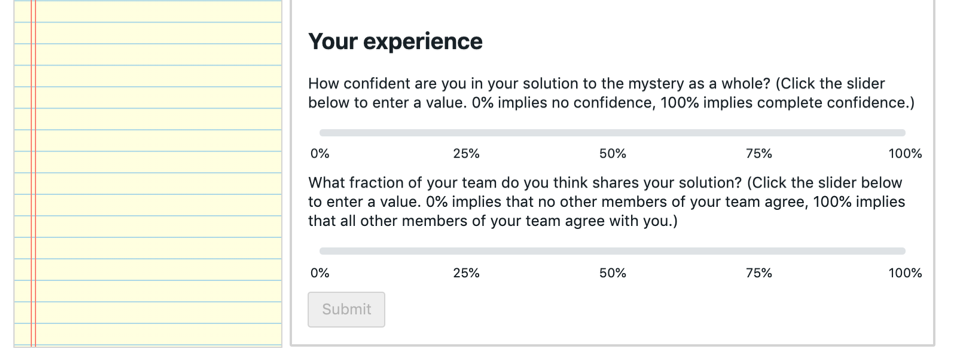}}
\caption{Post-game survey: Assess your confidence and estimate consensus}
\label{fig:confidence}
\end{figure}

Following the game, participants were asked to assess (using a slider) how likely it was that certain individuals referenced in the game were the burglar, and how likely it was that they used various tools, vehicles, and disguises in the task. The first few of these questions are shown in Fig. \ref{fig:make_the_case}. Sliders were labeled from Extremely Unlikely to Extremely Likely, and responses were recorded on a scale from 0 to 100. Participants were also asked to assess their confidence in their solution, and to estimate the level of consensus among their team, as shown in Fig. \ref{fig:confidence}.

After “Making the Case”, individuals were told that they were part of an experimental condition in which none of the clues were false, and they were rewarded \$0.10 for each clue in the promising leads section of their notebook, along with \$0.10 for each of the average number of clues their teammates had categorized as a promising lead. Participants were given a completion code to collect their bonuses, and an opportunity to report any problems with the game or describe their strategy. Many reported that they had enjoyed the game, and would like to play again.


\subsection{Experimental manipulation/stimulus}
\label{manipulation}
Clues were constructed from a bank of concepts (11 stolen objects, 11 crime scenes, 15 suspect names, 10 descriptions, 10 articles of clothing, 10 tools, and 10 vehicles) and a set of relationships (e.g. ``\{Name\} owns a \{vehicle\}'', ``A witness thought they saw \{stolen object\} in \{vehicle\}'') that formed a complete network between all concepts. This pool is sufficient to generate $11^2 * {15 \choose 3} * {10 \choose 2}^4 \approx 2.25\times10^{11}$ different mysteries.

\subsubsection{Constructing the bank of clue concepts}
\begin{figure}[h!]
\centering
\fbox{\includegraphics[width=0.7\columnwidth]{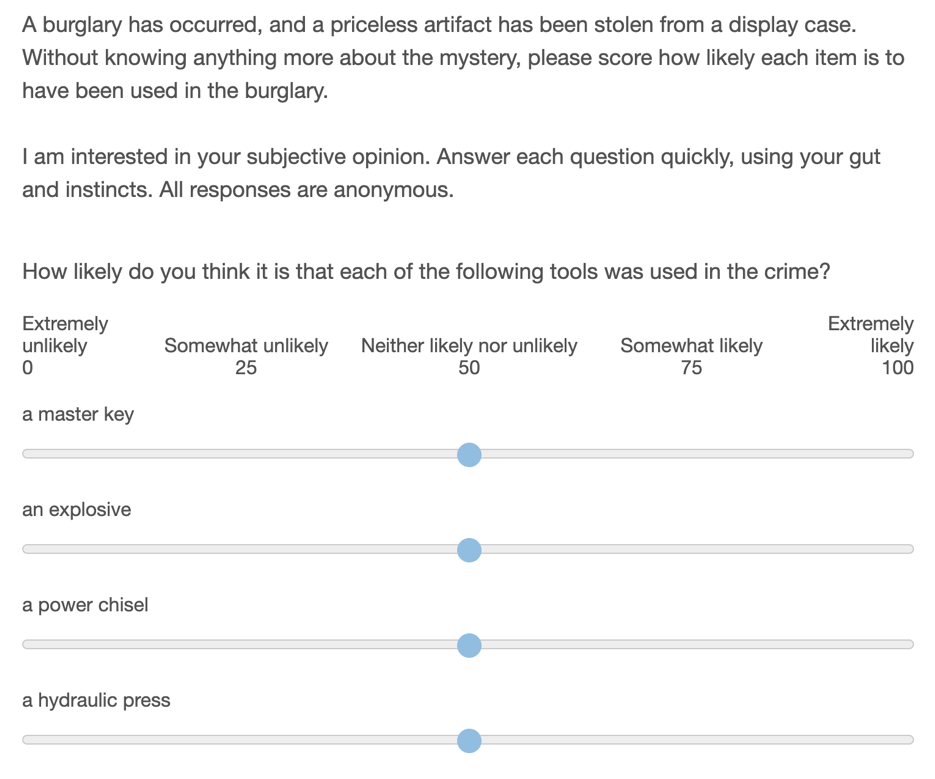}}
\caption{Pretesting the perceived likelihood of each concept being involved in a crime}
\label{fig:clue_pretest}
\end{figure}

The bank of concepts was constructed by starting with a pool of 403 candidate concepts including names, clothing, vehicles, etc. A pretest survey was conducted in which Amazon Mechanical Turk workers rated how likely each candidate concept was to have been used in a generic burglary. Individuals saw a subset of the concepts and were asked to give their gut reactions using a slider from Extremely Unlikely to Extremely Likely, as illustrated in Fig \ref{fig:clue_pretest}. In total, 139 participants rated each of the 403 candidate concepts between 20 and 30 times. Participants in the pretest were paid \$1.25 for a task which took each participant an average of about 4 minutes.

The pool of candidate names in the pretest represents the subset of the 200 most popular last names in the United States with a racial composition of between 50\% and 80\% ‘White’, as recorded in the 2000 US census. This selection is made to minimize the possibility of racial biases in the results. Additionally, names which are also common first names were excluded (e.g. “Stewart” or “Ross”) as were names which also serve as descriptors or adjectives in other clues (e.g. “Green”, “White”, or “Young”).

The remaining candidate concepts were written such that they would be as independent from one another as possible (e.g. I do not include both “a fat man” and “an overweight man” as these are synonymous, nor both “an old man” and “a man with grey hair” as these are perceived to go together.)

From the pretest results, I selected a subset of concepts that were perceived to be as likely as one another to be used in a burglary. (This helps to ensure that we do not see games in which all participants adopt “a set of lock picks” as a tool in the burglary, and reject “craft scissors”, just because lock picks are easier to imagine being used in a burglary.) The final selection was made by taking the subset of beliefs that minimized the difference in mean value of pretest survey responses when responses are normalized for each individual, and cross-checking against the means of the raw responses.

A similar pretest survey was conducted to select ‘spur’ clue concepts from a pool of candidates.

\newgeometry{margin=1cm} 
\begin{landscape}

\begin{table}[]
\caption {Clue concept bank. Each element is perceived to be equally likely to have been involved in a burglary.} \label{tab:elements}
\begin{tabular}{lllllll}
\textbf{CrimeScene}       & \textbf{StolenObject} & \textbf{Suspect} & \textbf{Clothing}        & \textbf{Appearance}   & \textbf{Tool}     & \textbf{Vehicle}          \\
the art museum            & the painting          & Collins          & a pair of overalls       & a long-haired man     & a hacksaw         & a yellow box truck        \\
the Pine Street Gallery   & the statue            & Hawkins          & a wool hat               & a pot-bellied man     & a serrated knife  & a blue Chevrolet Corvette \\
Kensington House          & the relic             & Mills            & a blue denim jacket      & a partially-bald man  & a set of hex keys & a green Mazda 3           \\
the Asper Casino          & the bracelet          & Cooper           & a tracksuit              & a grey-haired man     & a masonry drill   & a silver BMW              \\
the Danforth Hotel        & the antique           & Moore            & a pair of skinny-jeans   & a short man           & a circular saw    & a silver VW Jetta         \\
Knight Secure Storage     & the necklace          & Bennet           & a black scarf            & a well-groomed man    & a blowtorch       & a black Hummer            \\
the Daly Auction House    & the watch             & Mitchell         & a motorcycle helmet      & a man with sideburns  & an impact wrench  & a white Ford Fusion       \\
the Kentwood Mansion      & the diamond           & Stevens          & a black leather jacket   & a heavily-scarred man & a tire iron       & a blue Toyota Yaris       \\
the Dalhoff Estate        & the opal              & Wagner           & a pair of ripped jeans   & a blonde-haired man   & a pipe cutter     & a white Toyota Avalon     \\
the Darrowby Country Club & the crystal           & Edwards          & a blue long sleeve shirt & a handsome man        & a sledgehammer    & a blue Honda Fit          \\
DeRolfe Jewelers          & the jewel             & Rice             &                          &                       &                   &                           \\
                          &                       & Roberts          &                          &                       &                   &                           \\
                          &                       & Daniels          &                          &                       &                   &                           \\
                          &                       & Warren           &                          &                       &                   &                           \\
                          &                       & Sullivan         &                          &                       &                   &
\end{tabular}
\end{table}
\end{landscape}
\restoregeometry

\newgeometry{margin=1cm} 
\begin{landscape}
\begin{table}[]
\caption {Interdependent clue connections together create a complete knowledge graph.} \label{tab:inter_edges}
\begin{tabular}{ll}
\{StolenObject\_1\}   was kept in a case at \{CrimeScene\_1\} & \{Suspect\_2\} had been seen with   \{Tool\_2\} \\
\{Suspect\_1\}   was seen at \{CrimeScene\_1\} & \{Suspect\_2\} owns \{Vehicle\_1\} \\
\{Suspect\_2\}   was seen at \{CrimeScene\_1\} & \{Suspect\_2\} owns \{Vehicle\_2\} \\
\{Suspect\_3\}   was seen at \{CrimeScene\_1\} & \{Suspect\_3\} was known to wear   \{Clothing\_1\} \\
A person   wearing \{Clothing\_1\} was seen at \{CrimeScene\_1\} & \{Suspect\_3\} was known to wear   \{Clothing\_2\} \\
A person   wearing \{Clothing\_2\} was seen at \{CrimeScene\_1\} & \{Suspect\_3\} was described as   \{Appearance\_1\} \\
\{Appearance\_1\}   was seen at \{CrimeScene\_1\} & \{Suspect\_3\} was described as   \{Appearance\_2\} \\
\{Appearance\_2\}   was seen at \{CrimeScene\_1\} & \{Suspect\_3\} had been seen with   \{Tool\_1\} \\
Evidence   at \{CrimeScene\_1\} indicates the use of \{Tool\_1\} & \{Suspect\_3\} had been seen with   \{Tool\_2\} \\
Evidence   at \{CrimeScene\_1\} indicates the use of \{Tool\_2\} & \{Suspect\_3\} owns \{Vehicle\_1\} \\
\{Vehicle\_1\}   was seen leaving \{CrimeScene\_1\} & \{Suspect\_3\} owns \{Vehicle\_2\} \\
\{Vehicle\_2\}   was seen leaving \{CrimeScene\_1\} & \{Clothing\_1\} was found with   \{Clothing\_2\} \\
\{Suspect\_1\}   knew all about \{StolenObject\_1\} & \{Appearance\_1\} was seen wearing   \{Clothing\_1\} \\
\{Suspect\_2\}   knew all about \{StolenObject\_1\} & \{Appearance\_2\} was seen wearing   \{Clothing\_1\} \\
\{Suspect\_3\}   knew all about \{StolenObject\_1\} & \{Clothing\_1\} was found with   \{Tool\_1\} \\
A person   wearing \{Clothing\_1\} had been seen lurking near \{StolenObject\_1\} & \{Clothing\_1\} was found with   \{Tool\_2\} \\
A person   wearing \{Clothing\_2\} had been seen lurking near \{StolenObject\_1\} & \{Clothing\_1\} was found in   \{Vehicle\_1\} \\
\{Appearance\_1\}   had been lurking near \{StolenObject\_1\} & \{Clothing\_1\} was found in   \{Vehicle\_2\} \\
\{Appearance\_2\}   had been lurking near \{StolenObject\_1\} & \{Appearance\_1\} was seen wearing   \{Clothing\_2\} \\
The case   for \{StolenObject\_1\} might have been opened using \{Tool\_1\} & \{Appearance\_2\} was seen wearing   \{Clothing\_2\} \\
The case   for \{StolenObject\_1\} might have been opened using \{Tool\_2\} & \{Clothing\_2\} was found with   \{Tool\_1\} \\
A   witness thought they saw \{StolenObject\_1\} in \{Vehicle\_1\} & \{Clothing\_2\} was found with   \{Tool\_2\} \\
A   witness thought they saw \{StolenObject\_1\} in \{Vehicle\_2\} & \{Clothing\_2\} was found in   \{Vehicle\_1\} \\
\{Suspect\_2\}   hangs out with \{Suspect\_1\} & \{Clothing\_2\} was found in   \{Vehicle\_2\} \\
\{Suspect\_1\}   hangs out with \{Suspect\_3\} & \{Appearance\_2\} was seen with   \{Appearance\_1\} \\
\{Suspect\_1\}   was known to wear \{Clothing\_1\} & \{Appearance\_1\} was seen with   \{Tool\_1\} \\
\{Suspect\_1\}   was known to wear \{Clothing\_2\} & \{Appearance\_1\} was seen with   \{Tool\_2\} \\
\{Suspect\_1\}   was described as \{Appearance\_1\} & \{Appearance\_1\} was seen driving   \{Vehicle\_1\} \\
\{Suspect\_1\}   was described as \{Appearance\_2\} & \{Appearance\_1\} was seen driving   \{Vehicle\_2\} \\
\{Suspect\_1\}   had been seen with \{Tool\_1\} & \{Appearance\_2\} was seen with   \{Tool\_1\} \\
\{Suspect\_1\}   had been seen with \{Tool\_2\} & \{Appearance\_2\} was seen with   \{Tool\_2\} \\
\{Suspect\_1\}   owns \{Vehicle\_1\} & \{Appearance\_2\} was seen driving   \{Vehicle\_1\} \\
\{Suspect\_1\}   owns \{Vehicle\_2\} & \{Appearance\_2\} was seen driving   \{Vehicle\_2\} \\
\{Suspect\_3\}   hangs out with \{Suspect\_2\} & \{Tool\_1\} was found with \{Tool\_2\} \\
\{Suspect\_2\}   was known to wear \{Clothing\_1\} & \{Tool\_1\} was found in   \{Vehicle\_1\} \\
\{Suspect\_2\}   was known to wear \{Clothing\_2\} & \{Tool\_1\} was found in   \{Vehicle\_2\} \\
\{Suspect\_2\}   was described as \{Appearance\_1\} & \{Tool\_2\} was found in   \{Vehicle\_1\} \\
\{Suspect\_2\}   was described as \{Appearance\_2\} & \{Tool\_2\} was found in   \{Vehicle\_2\} \\
\{Suspect\_2\}   had been seen with \{Tool\_1\} & \{Vehicle\_2\} was found near   \{Vehicle\_1\}
\end{tabular}
\end{table}

\end{landscape}
\restoregeometry

\newgeometry{margin=1cm} 
\begin{landscape}

\begin{table}[]
\caption {Filler element bank. Each element equally influences the perception that an associated concept was involved in a burglary.} \label{tab:filler}
\begin{tabular}{lllll}
\textbf{suspectAge} & \textbf{suspectConviction} & \textbf{suspectMeans}                          & \textbf{suspectMotive}                    & \textbf{suspectTattoo} \\
33 years old        & drug possession            & was trained as a welder                        & has paid hush-money to a former lover     & a compass              \\
37 years old        & fraud                      & installs security systems                      & has family connections to organized crime & a bear                 \\
in their early 30's & drug distribution          & was trained as a goldsmith                     & is deep in payday-loan debt               & a heart                \\
in their mid 20's   & running a Ponzi scam       & has worked as an automotive repossession agent & has an expensive drug habit               & a flower               \\
in their late 20's  & embezzlement               & has worked as a security guard                 & has a gambling addiction                  & a clock                \\
in their late 30's  & identity theft             & worked at a pawn shop                          & wrote a revolutionary manifesto           & a star                 \\
in their early 20's & perjury                    & has worked as an armored car driver            & had been involved in gang activity        & a dog                  \\
29 years old        & shoplifting                & has worked for an import/export company        & has large gambling debts                  & a crown                \\
36 years old        & arson                      & worked for an alarm company                    & has a heroin addiction                    & a flag
\end{tabular}
\end{table}

\begin{table}[]
\begin{tabular}{lllll}
\textbf{appearanceInjury} & \textbf{appearanceRemoved} & \textbf{appearanceReported}   & \textbf{appearanceStreet} & \textbf{appearanceWanted}                         \\
a broken arm              & a museum                   & waiting in a dark alley       & Maple Avenue              & Law enforcement is seeking                        \\
a fractured kneecap       & a public library           & shouting at 11pm              & Lincoln Boulevard         & Officers are asking questions about               \\
a concussion              & a bar                      & sitting in a tree             & Chestnut Street           & Private security companies have been warned to... \\
a fractured rib           & a restaurant               & painting graffiti             & Church Street             & Airport security has been asked to look out for   \\
minor burns               & a residence                & vandalizing a vending machine & Hill Street               & Police are interviewing witnesses about           \\
a drug overdose           & a party                    & carrying a large bag          & Ninth Avenue              & Information is wanted about
\end{tabular}
\end{table}

\begin{table}[]
\begin{tabular}{lllll}
\textbf{carBehavior}                     & \textbf{carBuy}          & \textbf{carDamage}    & \textbf{carEnterprise} & \textbf{carTicketed}       \\
driving after midnight                   & in a wholesale auction   & a broken headlight    & a club                 & an expired registration    \\
with someone sleeping in the   back seat & at a police auction      & a broken grill        & a massage parlor       & parking in a loading zone  \\
with darkly tinted windows               & from a classified ad     & damaged suspension    & a strip mall           & driving without headlights \\
parked in a lot for multiple   nights    & at an estate sale        & a missing wing mirror & a laundromat           & a broken tail light        \\
taking the back streets                  & from a used-car salesmen & the airbags deployed  & a delicatessen         & illegal parking            \\
with its hood up on the   roadside       & from a junk-yard         & a broken axle         & a hotel                & running a stop sign
\end{tabular}
\end{table}

\begin{table}[]
\begin{tabular}{lllll}
\textbf{clothingActivity}             & \textbf{clothingDamage} & \textbf{clothingDiscoverer}                  & \textbf{clothingFootage} & \textbf{clothingWith}                  \\
pacing back and forth                 & cut into pieces         & a gym owner emptying abandoned lockers       & at a bus stop            & a list of tools                        \\
entering a machine room               & with tire marks on it   & a dockworker moving shipping pallets         & in the woods             & a home-made electronic device          \\
pulling an object out of a   gutter   & with frayed edges       & a store worker breaking down cardboard boxes & in the park              & a pair of rubber gloves                \\
looking through binoculars at   night & burned in a fire        & a journalist uncovering a story              & at a campsite            & an inter-city train schedule           \\
getting into a taxi                   & discolored with bleach  & a postal worker emptying a mailbox           & on a bridge              & the stub of a bus ticket               \\
climbing on a bridge                  & caked in mud            & theater staff cleaning up after a movie      & on the golf course       & an envelope containing GPS coordinates
\end{tabular}
\end{table}

\begin{table}[]
\begin{tabular}{lllll}
\textbf{toolDamage}                         & \textbf{toolFound}    & \textbf{toolUse}               & \textbf{toolWith}                  & \textbf{toolrandom}                            \\
showing signs of misuse                     & buried in debris      & access maintenance crawlspaces & with a pair of work gloves         & could be used by one person                    \\
with minor damage                           & in an abandoned house & open an upper-story window     & with safety features removed       & has been used in prior burglaries \\
with burn marks                             & in a trash compactor  & bypass an alarm system         & that had been painted black        & could be concealed in a backpack               \\
disassembled into pieces & in a garage           & deactivate a motion sensor     & with gunpowder residue             & leaves distinctive marks if used carelessly    \\
covered in sawdust  & in a creek            & disassemble an alarm panel     & wrapped in newspaper               & is often used by thieves                       \\
\begin{minipage}[t]{0.15\columnwidth}that had been damaged falling from a height\end{minipage}                    & beside a road         & circumvent a lock              & wrapped in tape to make it quieter & \begin{minipage}[t]{0.25\columnwidth}was shown in news coverage of another burglary  \end{minipage}
\end{tabular}
\end{table}

\end{landscape}
\restoregeometry

\newgeometry{margin=1cm} 
\begin{landscape}

\begin{table}[]
\caption {Independent clue connections break relationships between analysis clues} \label{tab:indep_edges}
\begin{tabular}{ll}
\{StolenObject\_1\}   was kept in a case at \{CrimeScene\_1\} & A constable noticed   \{Appearance\_1\} on \{appearanceStreet\_1\} \\
\{Suspect\_1\}   was seen at \{CrimeScene\_1\} & A constable noticed   \{Appearance\_2\} on \{appearanceStreet\_2\} \\
\{Susect\_2\}   was seen at \{CrimeScene\_1\} & \{Tool\_1\} was found \{toolWith\_1\} \\
\{Suspect\_3\}   was seen at \{CrimeScene\_1\} & \{Tool\_2\} was found \{toolWith\_2\} \\
A person   wearing \{Clothing\_1\} was seen at \{CrimeScene\_1\} & \{Vehicle\_1\} was recently   purchased \{carBuy\_1\} \\
A person   wearing \{Clothing\_2\} was seen at \{CrimeScene\_1\} & \{Vehicle\_2\} was recently   purchased \{carBuy\_2\} \\
\{Appearance\_1\}   was seen at \{CrimeScene\_1\} & \{Suspect\_1\} is \{suspectAge\_1\} \\
\{Appearance\_2\}   was seen at \{CrimeScene\_1\} & \{Suspect\_2\} is \{suspectAge\_2\} \\
Evidence   at \{CrimeScene\_1\} indicates the use of \{Tool\_1\} & \{Suspect\_3\} is \{suspectAge\_3\} \\
Evidence   at \{CrimeScene\_1\} indicates the use of \{Tool\_2\} & \{Clothing\_1\} was discovered by   \{clothingDiscoverer\_1\} \\
\{Vehicle\_1\}   was seen leaving \{CrimeScene\_1\} & \{Clothing\_2\} was discovered by   \{clothingDiscoverer\_2\} \\
\{Vehicle\_2\}   was seen leaving \{CrimeScene\_1\} & \{Appearance\_1\} was reported   \{appearanceReported\_1\} \\
\{Suspect\_1\}   knew all about \{StolenObject\_1\} & \{Appearance\_2\} was reported   \{appearanceReported\_2\} \\
\{Suspect\_2\}   knew all about \{StolenObject\_1\} & \{Tool\_1\} could be used to   \{toolUse\_1\} \\
\{Suspect\_3\}   knew all about \{StolenObject\_1\} & \{Tool\_2\} could be used to   \{toolUse\_2\} \\
A person   wearing \{Clothing\_1\} had been seen lurking near \{StolenObject\_1\} & \{Vehicle\_1\} was ticketed for   \{carTicketed\_1\} \\
A person   wearing \{Clothing\_2\} had been seen lurking near \{StolenObject\_1\} & \{Vehicle\_2\} was ticketed for   \{carTicketed\_2\} \\
\{Appearance\_1\}   had been lurking near \{StolenObject\_1\} & \{Suspect\_1\} \{suspectMeans\_1\} \\
\{Appearance\_2\}   had been lurking near \{StolenObject\_1\} & \{Suspect\_2\} \{suspectMeans\_2\} \\
The case   for \{StolenObject\_1\} might have been opened using \{Tool\_1\} & \{Suspect\_3\} \{suspectMeans\_3\} \\
The case   for \{StolenObject\_1\} might have been opened using \{Tool\_2\} & \{Clothing\_1\} was found with   \{clothingWith\_1\} \\
A   witness thought they saw \{StolenObject\_1\} in \{Vehicle\_1\} & \{Clothing\_2\} was found with   \{clothingWith\_2\} \\
A   witness thought they saw \{StolenObject\_1\} in \{Vehicle\_2\} & \{Appearance\_1\} was treated for   \{appearanceInjury\_1\} \\
\{Suspect\_1\}   has a tattoo of \{suspectTattoo\_1\} & \{Appearance\_2\} was treated for   \{appearanceInjury\_2\} \\
\{Suspect\_2\}   has a tattoo of \{suspectTattoo\_2\} & An FBI agent found \{Tool\_1\}   \{toolFound\_1\} \\
\{Suspect\_3\}   has a tattoo of \{suspectTattoo\_3\} & An FBI agent found \{Tool\_2\}   \{toolFound\_2\} \\
A   policeman saw someone in \{Clothing\_1\} \{clothingActivity\_1\} & An officer identified   \{Vehicle\_1\} at \{carEnterprise\_1\} \\
A   policeman saw someone in \{Clothing\_2\} \{clothingActivity\_2\} & An officer identified   \{Vehicle\_2\} at \{carEnterprise\_2\} \\
\{appearanceWanted\_1\}   \{Appearance\_1\} & \{Suspect\_1\} has a prior   conviction for \{suspectConviction\_1\} \\
\{appearanceWanted\_2\}   \{Appearance\_2\} & \{Suspect\_2\} has a prior   conviction for \{suspectConviction\_2\} \\
\{Tool\_1\}   \{toolrandom\_1\} & \{Suspect\_3\} has a prior   conviction for \{suspectConviction\_3\} \\
\{Tool\_2\}   \{toolrandom\_2\} & Forensics identified   \{Clothing\_1\} \{clothingDamage\_1\} \\
\{Vehicle\_1\}   was reported \{carBehavior\_1\} & Forensics identified   \{Clothing\_2\} \{clothingDamage\_2\} \\
\{Vehicle\_2\}   was reported \{carBehavior\_2\} & \{Appearance\_1\} was forcibly   removed from \{appearanceRemoved\_1\} \\
\{Suspect\_1\}   \{suspectMotive\_1\} & \{Appearance\_2\} was forcibly   removed from \{appearanceRemoved\_2\} \\
\{Suspect\_2\}   \{suspectMotive\_2\} & A forensics report contained   \{Tool\_1\} \{toolDamage\_1\} \\
\{Suspect\_3\}   \{suspectMotive\_3\} & A forensics report contained   \{Tool\_2\} \{toolDamage\_2\} \\
Someone   wearing \{Clothing\_1\} was seen on security footage \{clothingFootage\_1\} & \{Vehicle\_1\} was found with   \{carDamage\_1\} \\
Someone   wearing \{Clothing\_2\} was seen on security footage \{clothingFootage\_2\} & \{Vehicle\_2\} was found with   \{carDamage\_2\}
\end{tabular}
\end{table}

\end{landscape}
\restoregeometry

\subsubsection{Assembling sets of clues for use in games}
This experiment manipulates the structure of clues within the detective game, to create an ``interdependent'' condition in which the clues interact strongly with each other, and an ``independent'' condition that limits those interactions while preserving as much similarity with the treatment condition as possible.

\begin{figure}[h!]
\centering
\includegraphics[width=0.7\columnwidth]{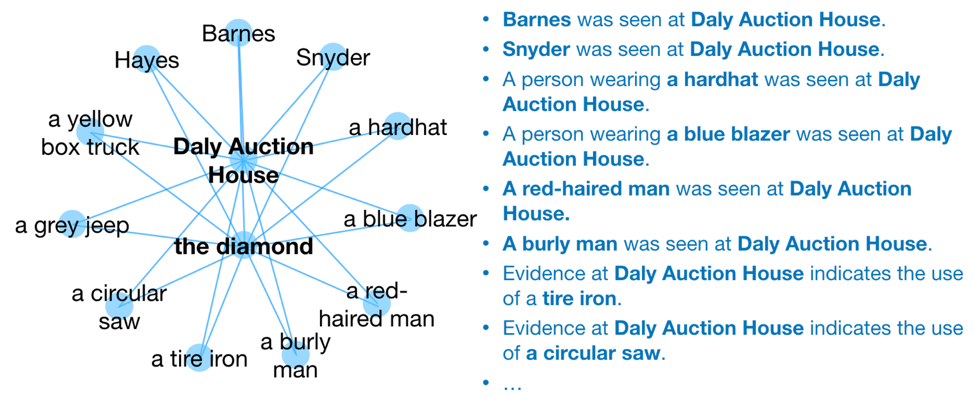}
\caption{Wave 1 of clue generation created ``spoke'' or analysis clues used in both treatment and control games. ``Spoke'' clues connected rim concepts to hub concepts.}
\label{fig:spokes}
\end{figure}

Clues were constructed in three waves. The first wave was identical for both conditions, and is illustrated in Fig. \ref{fig:spokes}. Clues were created that link ‘hub’ concepts (including a crime scene and a stolen object) to ‘rim’ concepts (including three suspects, two articles of clothing, two physical descriptions, two tools, and two vehicles). For example “\textbf{Hayes} was seen at the \textbf{Daly Auction House}” or “The case for the \textbf{diamond} might have been opened using a \textbf{circular saw}”. “Spoke” clues were independent of one another, as they could only interact via association with the crime scene and stolen object – items that were known in advance to be relevant to the mystery. There were 11 rim concepts and 2 hub concepts, and so 22 spoke clues.

\begin{figure}[h!]
\centering
\includegraphics[width=0.7\columnwidth]{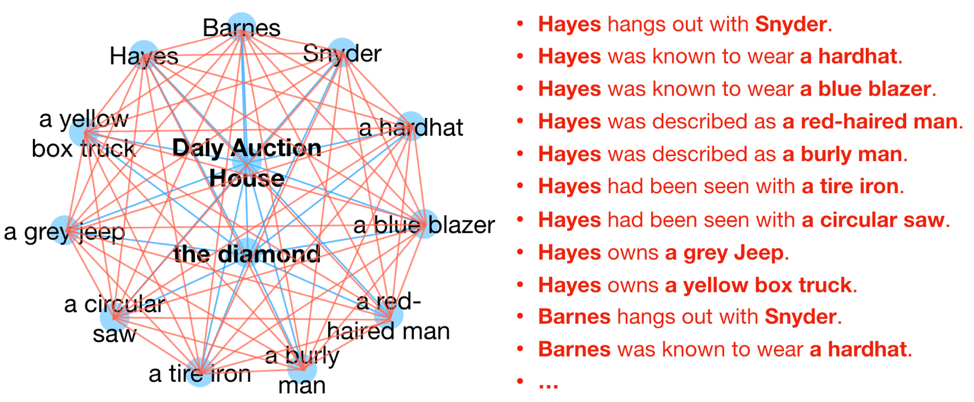}
\caption{Wave 2 of clue generation created ``cross-link'' clues for the interdependent condition. Cross-link clues connect rim concepts to one another.}
\label{fig:crosslinks}
\end{figure}

In the interdependent condition, the second wave of clue construction created ``cross-link'' clues, which connected each of the spoke clues to one another (e.g. ``\textbf{Hayes} owns a \textbf{circular saw}''). These cross-link clues create interdependence between the spoke clues, and allow for clues to logically support one another (e.g. if I believe that ``A burly man was seen at the Daly Auction House'' and that ``Barnes is a burly man'', then I am more receptive to the idea that ``Barnes was seen at the Daly Auction House''). A cross-link clue connected each of the 11 rim concepts to the other rim concepts, for a total of 55 unique cross-link clues, as shown in Fig. \ref{fig:crosslinks}.

\begin{figure}[h!]
\centering
\includegraphics[width=0.7\columnwidth]{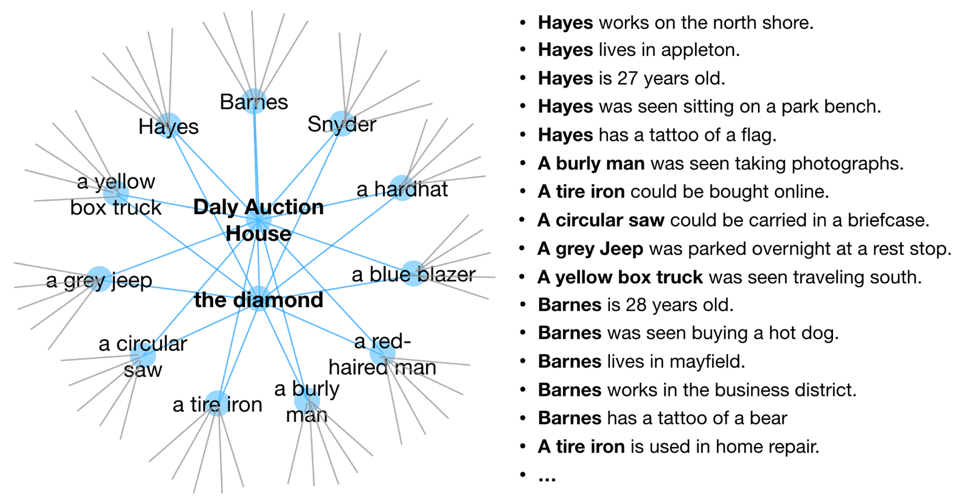}
\caption{Wave 2 of clue generation created ``spur'' clues for the independent condition. Spur clues filled the place of cross-link clues without creating links between rim concepts. Spur clues still allowing for multiple exposures to rim concepts.}
\label{fig:spurs}
\end{figure}

In the independent condition, the second wave of clue construction created “spur” clues that connected to the rim concepts, but did not connect to other clues (Fig \ref{fig:spurs}). There were the same number of spur clues in the independent condition as cross-link clues in the interdependent condition: 55. By connecting to the rim concepts (rather than being disconnected altogether) these clues help separate the effect of interdependence manifest as logical relationships between clues from the effect of the frequency of each rim concept in the set of clues. The content of the spur clues was selected in pretest to have a uniform impact on participants judgement of the rim element to which they connect.

The first and second waves of clue construction created 77 unique clues. As there were 20 individuals in each treatment within each game, 80 clues were needed to give each individual 4 starting clues. The third wave of clue construction filled the 3 remaining spaces with the clue connecting the crime scene to the stolen object (e.g. The \textbf{diamond} was stolen from the \textbf{Daly Auction House}.) This was redundant information, as all participants were told this at the start of the game.

\subsection{Data collection and measurements}
\label{data_collection}
Sufficient data was collected to replicate the state of the game at any point during game play, and to observe every action taken by every player.

\subsubsection{Recording player actions}
In addition to the social network structure and the initial assignment of clues to positions within the social network, the following information was recorded:
\begin{enumerate}
    \item Each drag event that resulted in a change in a player’s notebook (i.e. addition of a clue to a notebook section OR change of order within a notebook section) was logged. Logging information includes the ID of the clue being dropped, the source for the drag event (i.e. the exposing player or notebook the belief came from), the destination for the drag event (i.e. the notebook the clue is being dragged into), the position within the destination notebook that the clue moved into (i.e. its index in the list) and the time at which the drop event occurred.
    \item The final state of all detectives' notebooks was recorded. Together with the initial state, this provides a check that all events were logged properly.
    \item Each individual provided a self-report of the degree to which they believed each of the 11 “rim” concepts to be connected to the crime, collected using an empty slider from “Extremely Unlikely” to “Extremely Likely”. Slider positions were captured as an integer value between 0 and 100.
    \item Individuals reported their confidence in their solution on a scale from 0 to 100 using a blank slider.
    \item Individuals reported the fraction of their team they thought shared their solution on a scale from 0 to 100 percent, using a blank slider.
    \item Individuals reported their Age, Education, Gender, and feedback on the game.
\end{enumerate}

\subsubsection{Choice of summary statistics}

The \textbf{similarity between individuals' self-reported beliefs} was measured using Pearson's correlation coefficient.
As others have identified, \cite{goldberg2018beyond} correlation is a natural measure when we have a fixed number of continuous measures of each subject. It is readily interpretable, and the fixed range (-1,1) maps to intuitive understandings of similarity and difference.

The \textbf{similarity between individuals' behavior} was measured using the Phi coefficient. This measure corresponds to Pearson correlation when values are binary, and has the same interpretable (-1,1) range.
The phi coefficient is appropriate for a universe with a finite number of beliefs, but would be less appropriate as the number of adopted beliefs becomes a very small fraction of the total number of possible beliefs.

The \textbf{alignment of the population along a ``left-right axis''} was measured as the percent of variance present in the first principal component, using singular value decomposition.
This measure corresponds to the notion of “constraint” articulated by Dimaggio et al. (\cite{dimaggio1996have}). In their paper they describe Chronbach’s alpha and the PCA measure both providing similar measures of constraint. I have chosen the PCA measure here as more interpretable, as it maps better to our intuitive understanding of a political spectrum.

The \textbf{within-camp and across-camp similarities} were measured as the 5\textsuperscript{th} and 95\textsuperscript{th} percentile similarities according to the similarity measures above.
There are a number of different measures in the literature that try to capture the notion that with polarization, the most similar individuals become more self-similar, and the least similar individuals move further away from one another.
The fact that no single measure has emerged as the leader hints at problems with each.
When the identities of camps are already known the difference of means between groups can be used (\cite{becker2019wisdom},\cite{dimaggio1996have}).
Variance (see \cite{baldassarri2007dynamics},\cite{dimaggio1996have}) captures heterogeneity between individuals, but not clustering into camps.
Kurtosis (\cite{baldassarri2007dynamics},\cite{dimaggio1996have}) is predicated on a bimodal distribution.
The “gap” statistic (\cite{goldberg2018beyond}) is one of dozens of ways of assessing the quality of a machine learning clustering algorithm.

As I do not need to identify the groups themselves, or compare to external datasets, it is sufficient to merely report what each of these other measures is trying to approximate: the similarity that is found within groups, and that which is found across groups. As I am only interested in the relative differences between conditions (or changes over time) then I can arbitrarily designate a threshold for which comparisons will be considered ``within-group'' or ``across groups''. This provides a much more intuitive demonstration of increasing polarization than the measures found in above literature.

The closer the chosen thresholds are to the tails of the distribution, the more conservative the claim that the comparisons beyond this threshold are appropriately ``within'' or ``across'' groups. At the same time, we need enough samples included in the set to minimize noise due to the finite number of comparisons. In this 20-participant social network, the 95\textsuperscript{th} and 5\textsuperscript{th} percentiles correspond to 10 comparisons between individuals. The sensitivity of results to the choice of these thresholds is explored in section \ref{section:thresholds} of this supplement.

For behavioral measures of within and across-camp similarity (i.e. those based upon clues identified as promising leads at the end of the game) these percentiles are sensitive not only to interdependence and network structure, but also to the average level of diffusion of clues. As the average level of diffusion can vary between games due to differences in players' baseline activity levels, this additional source of noise increases the sample size we would require to see an effect. To compensate for this noise, we can compare the percentiles to what would be expected due to chance, given the same extent of diffusion. To calculate the effect of interdependence and network structure, I first remove some of the noise by subtracting the 5\textsuperscript{th} and 95\textsuperscript{th} percentile values from a shuffled data-set that keeps the number of adopters of each clue and the number of clues adopted by each participant fixed. This makes for an apples-to-apples comparison of the effects of interdependence or network structure. This correction was designed in simulation and included in the preregistration for the experiment.

\subsubsection{Handling missing data}

As the game was played in real-time, the effect of a participant ‘dropping out’ during game-play was equivalent to them holding their beliefs fixed for the remainder of the game. As it is impossible to distinguish these two behaviors, I identified a drop-out as any player failing to submit the post-game survey.

When an individual failed to complete the post-game survey, aggregate results for their condition were calculated based upon the remaining players. Aggregate results for the paired comparison conditions were calculated as the average of all same-sized subsets of players in each comparison condition.

\subsection{Preregistration}
\label{prereg}

The preregistration for the experiment included all the code necessary to run and analyze the experiment presented in this paper, such that a direct replication can be conducted from the preregistered materials. The preregistration is available at \href{https://osf.io/239ns}{https://osf.io/239ns}. In addition to the analyses presented in this paper, the preregistration also included a secondary analysis of mediating and moderating effects of interdependent diffusion that will be reported in a second paper.

The experiment differed from the preregistered procedure in three minor ways: expanding the recruitment pool, correcting a coding mistake, and formalizing an analysis that was preregistered with only a text description.

\begin{enumerate}
    \item The preregistered strategy of recruiting individuals with a “sign-up” HIT immediately prior to each game did not scale well to multiple games per day. Instead, I drew on a pool of Mechanical Turk workers from the US and Canada that had been recruited for a previous experiment. As the locations of the panelists were not recorded, this required me to relax the preregistered participant qualifications to include Canadian workers. I felt that this addition would not significantly influence the likelihood of success of the experiment or it’s generalizability. The phenomenon under study is not expected to behave differently in different populations, and the study does not require any special outside knowledge.

    \item The analysis code for comparing end-of-game measures was designed originally to account for dropouts in a (treatment/control) game by averaging over same-sized subsets of the matching (control/treatment) game. However, the code as preregistered did not account for the fact that comparisons should be made between four experimental conditions instead of two. The code was revised to make the correct comparisons.

    \item While the interaction analysis was included in the text of the preregistration, the code for computing it was mistakenly omitted. This additional code is included in the experiment repository.

\end{enumerate}

\subsection{Outcomes}
\label{outcomes}

\subsubsection{Participants' experience}

Participants were Amazon Mechanical Turk workers who lived in the US or Canada and were over 18 years of age. Workers must have completed at least 100 HITs and have a 90\% or better approval rating. Recruitment and compensation were handled using TurkPrime (\url{www.cloudresearch.com}).

For blocks 0-2, recruitment took place in the hour preceding a game launch via a timed “sign-up” HIT. This recruitment strategy had been shown to be effective during pilots, but did not scale well when multiple blocks were run during the same day. For blocks 3-29, recruitment took advantage of a panel of workers who had previously indicated willingness to be notified of upcoming games. This panel was expanded through ongoing paid and unpaid recruitment HITs during the period the experiments were run. Panel members were notified via email at the beginning of each day when experiments would take place, and again 10 minutes prior to the launch of a game. At the scheduled game time, HITs were made available to any worker meeting the qualifications, whether they were in the original panel or not.

Participants were compensated \$0.10 for accepting the game HIT, in addition to \$1 for training, \$1 for playing the game, and up to \$2 in bonuses. Participants who trained but were unable to play were eligible to attempt to play again. Those who completed training and entered the game were blocked from participating in future games via an exclusion qualification.

\begin{figure}[H]
\centering
\includegraphics[width=0.6\columnwidth]{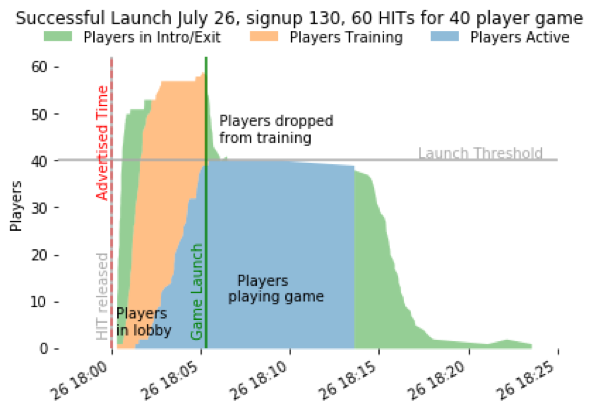}
\caption{Players active by stage and time. (Data from pilot test.)}
\label{fig:recruitment}
\end{figure}

The game took about 20 minutes to play, including training, waiting room, and follow-up. The average payout was approximately \$4.00, for an hourly rate of approximately \$12.00/hr. Participants who trained but were unable to play spent about 5 minutes in the platform before they were bumped. They earned \$1.10, for an approximate hourly rate of \$13.20.  Fig. \ref{fig:recruitment} shows the number of participants active in different parts of the game at different times for a 2-condition pilot test. There were necessarily some individuals who were dropped from training when the game launched, as the number of individuals who would showed up for a game was unpredictable.

\subsubsection{Detailed Results}
Over eight minutes of gameplay, participants on average made 28.4 classifications and adopted 16.2 clues as promising leads.
Despite the symmetry of the setup, and the fact that there was by design no solution to the mystery, participants came to strongly-held beliefs about which suspect was guilty and how they performed the crime. On average, participants felt that the most likely suspect was almost 50\% more likely to be involved in the crime than the least likely suspect, and over half of participants reported confidence in at least one part of their solution of 95\% or greater. Table \ref{Tab:effect_summary} reports the effect sizes for all comparisons made in the experiment.

\begin{table}[h]
\centering
\caption{Effect and Interaction Summary}
\label{Tab:effect_summary}

\begin{tabular}{llcccccc}
\hline
 & \multicolumn{1}{l|}{}              & \multicolumn{3}{c|}{\textbf{Effect of Interdependence}}        & \multicolumn{3}{c}{\textbf{Effect of Network Structure}} \\ \cline{3-8}
 & \multicolumn{1}{l|}{}              & Effect Size & p value & \multicolumn{1}{c|}{90\% CI}           & Effect Size       & p-value       & 90\% CI              \\ \hline
\multirow{2}{*}{\begin{tabular}[c]{@{}l@{}}Within-Camp\\ Similarity\end{tabular}} &
  \multicolumn{1}{l|}{- Behavioral} &
  0.026 &
  0.015 &
  \multicolumn{1}{c|}{(0.0083,  0.045)} &
  .16 &
  1.17e-9 &
  (.13, .19) \\
 & \multicolumn{1}{l|}{- Self-Report} & -0.0039     & 0.41    & \multicolumn{1}{c|}{(-0.036,  0.025)}  & .021              & .12           & (.0099, .050)        \\ \hline
\multirow{2}{*}{\begin{tabular}[c]{@{}l@{}}Across-Camp\\ Similarity\end{tabular}} &
  \multicolumn{1}{l|}{- Behavioral} &
  0.010 &
  0.22 &
  \multicolumn{1}{c|}{(-0.012,  0.032)} &
  -.034 &
  .004 &
  (-.054, -.016) \\
 & \multicolumn{1}{l|}{- Self-Report} & -0.034      & 0.058   & \multicolumn{1}{c|}{(-0.067, 0.0015)}  & -.089             & 6.3e-5        & (-.12, -.057)        \\ \hline
\multirow{2}{*}{\begin{tabular}[c]{@{}l@{}}Alignment w/\\ Left-Right Axis\end{tabular}} &
  \multicolumn{1}{l|}{- Behavioral} &
  2.13\% &
  0.016 &
  \multicolumn{1}{c|}{(0.61\%, 3.72\%)} &
  13.7\% &
  2.1e-11 &
  (.12, .16) \\
 & \multicolumn{1}{l|}{- Self-Report} & 2.81\%        & 0.022   & \multicolumn{1}{c|}{(0.57\%,  5.03\%)} & 7.2\%             & 2.3e-5        & (4.8\%, 9.8\%)       \\ \hline
\multicolumn{8}{l}{\begin{tabular}[c]{@{}l@{}}Baseline for effect: Independent clues and non-polarizing social network. \\ One-sided pairwise T-tests; n=30 pairs; unadjusted p-values\end{tabular}} \\ \hline
\end{tabular}
\end{table}

The presented theory does not predict how individuals will feel about their team’s performance. Interdependence did not change the perceived consensus among the team by any measurable amount, despite the increase in polarization. However, participants in the interdependent condition did report feeling more confident in their solution (+2.17\% p=.045). This is not a particularly large effect, as the results were measured on a 100 point scale.

\section{Resources for replication, extension and reanalysis}

\subsection{Conditions of validity}
The simulations and effects presented here are only valid when susceptibility to a belief can vary over the same timescale as the diffusion process. This primarily occurs when there are multiple beliefs diffusing in the same social network over the same timescales. If the population is broadly susceptible to a belief before it becomes available for adoption (for example, if international relations are strained, news of war might propagate quickly, as individuals are already susceptible to this idea) then diffusion of that belief will proceed much like the spread of a viral infection. Other beliefs would certainly be spreading at the same time, but they would not be necessary for the adoption of the focal belief.

Conversely, if there are not enough facilitating beliefs in a population to make adoption likely, then the reciprocal facilitation mechanism is unlikely to activate, and diffusion (if it occurs at all) will be merely among those who are initially susceptible.

We can see both of these limiting conditions in simulation by changing the number of beliefs that an individual starts with. Too few, and diffusion stalls in both independent and interdependent simulations. Too many, and susceptibility is a foregone conclusion, and in both independent and interdependent conditions adoption is universal. However, for a broad range of values in between, reciprocal facilitation is the dominant factor in a particular belief's level of adoption.

The mechanisms presented in this paper are also likely to be less important when information spreads from a central source to all individuals, as the agreement cascade mechanism depends upon individuals throughout the network adopting beliefs from one another.

\subsection{Replicating these results}

All of the code required to conduct this experiment and all of the data generated by the experiment is available open-source at \href{https://github.com/JamesPHoughton/interdependent-diffusion}{https://github.com/JamesPHoughton/interdependent-diffusion}. An exact replication of these results can be run without writing any code. Slight changes to the replication - such as creating new sets of clues, or using a different social network - can be accomplished by changing the experiment's configuration file.

Resources that will help a researcher replicate this analysis include the Empirica documentation (\href{https://empirica.ly/}{https://empirica.ly/}) and introductory paper \cite{almaatouq2021empirica}, and the Empirica code repository: \href{https://github.com/empiricaly/meteor-empirica-core}{https://github.com/empiricaly/meteor-empirica-core}.

I am also happy to advise replication efforts and answer questions about the code or implementation at the github repository for this experiment: \href{https://github.com/JamesPHoughton/interdependent-diffusion/issues}{Repository Issues}.

\subsection{Extending this research}
There are a number of obvious opportunities for extending this research. In this simulation and experiment, individuals do not verify their beliefs against ground-truth. We should hope that in the real world, this occurs at least occasionally, and that when it does, it forms a corrective force against the drivers of polarization. An interesting extension to this experiment would be to test the effects of belief verification on macro-scale outcomes. An experimenter could manipulate the cost to verify information, and the correlation of this cost between clues, or between members of the population, and examine the effects of specialization or general knowledge on collective problem solving.

Another opportunity for extension would be to vary the way information is presented, to mirror that of various social media websites, and to explore which factors tend to amplify the polarizing influence of agreement cascades and reciprocal facilitation.

The detective game experiment allows researchers to study the simultaneous contagion of multiple diffusants on a level playing field; the game could be adapted to many contexts with this requirement.

Resources for extending this research include the Meteor (\url{https://www.meteor.com/}) and React (\url{https://reactjs.org/}) documentation, and the above-listed Empirica documentation. Additionally, several third-party developers have experience developing experiments using Empricia and can be engaged to implement modifications.

\subsection{Opportunities for reuse of the data generated by this experiment}
This experiment recorded timestamped data on 68,229 adoption decisions made by 2400 individuals, along with instantaneous and historical information participants used to make those decisions. The information originally shown to participants is randomized, and diverse. As a result, there are many opportunities to reuse this data to answer questions about the micro-level processes involved in the adoption of new beliefs.

\bibliographystyle{plain}
\bibliography{references}